\documentclass[aps,preprint,11pt]{revtex4}
\usepackage{amsmath,amssymb,amsthm,color}
\usepackage{epsfig}
\usepackage{graphicx}
\usepackage{amssymb}
\usepackage{amsfonts}
\usepackage{amsmath}
\usepackage{graphicx}%
\usepackage{subfigure}
\usepackage{setspace}

\newcommand{\SSS}[3]{\Sigma_{#1}\otimes\Sigma_{#2}\otimes\Sigma_{#3}}
\newcommand{\ISS}[2]{I\otimes\Sigma_{#1}\otimes\Sigma_{#2}}
\newcommand{\SIS}[2]{\Sigma_{#1}\otimes I\otimes\Sigma_{#2}}
\newcommand{\SSI}[2]{\Sigma_{#1}\otimes\Sigma_{#2}\otimes I}
\newcommand{\CC}[2]{\Sigma_{#1}\otimes\Sigma_{#2}}
\newcommand{\be}{\begin{equation}}
\newcommand{\ee}{\end{equation}}
\newcommand{\no}{\nonumber}
\newcommand{\ket}[1]{|#1\rangle}
\newcommand{\bra}[1]{\langle #1|}

\def\be#1{\begin{equation}\label{#1}}
\def\ee{\end{equation}}
\def\bea#1{\begin{eqnarray}\label{#1}}
\def\eea{\end{eqnarray}}

\def\r{\right)}

\begin{document}
\begin{spacing}{1.2}

\title{Tripartite Entanglements in Non-inertial Frames}

\author{M. Shamirzai}\email{mshj@iaush.ac.ir}\affiliation{Department of Physics, Faculty of
Sciences, University of Isfahan , Isfahan 81744, Iran}
\author{B. Nasr Esfahani }\email{ba_nasre@sci.ui.ac.ir}\affiliation{Department of Physics, Faculty of
Sciences, University of Isfahan , Isfahan 81744, Iran}
\author{M. Soltani}\email{msoltani@phys.ui.ac.ir}\affiliation{Department of Physics, Faculty of
Sciences, University of Isfahan , Isfahan 81744, Iran}

\begin{abstract}

Entanglement degradation caused by the Unruh effect is discussed
for the tripartite GHZ or W states constructed by modes of a
non-interacting quantum field viewed by one inertial observer and
two uniformly accelerated observers. For fermionic states, the
Unruh effect even for infinite accelerations cannot completely
remove the entanglement. However, for the bosonic states, the
situation is different and the entanglement vanishes
asymptotically. Also, the entanglement is studied for the
bipartite subsystems. While for the GHZ states all the bipartite
subsystems are identically disentangled, for the W states the
bipartite subsystems are somewhat entangled, though, this
entanglement can be removed for appropriately accelerated
observers. Interestingly, logarithmic negativity as a measure for
determining the entanglement of one part of the system relative to
the other two parts, is not generally the same for different
parts. This means that we encounter tripartite systems where each
part is differently entangled to the other two parts.
\end{abstract}

\maketitle

\section{Introduction}

The phenomenon of entanglement has a central importance in the
quantum information science, and has emerged as a fundamental
resource in quantum communication, quantum cryptography, quantum
teleportation and quantum computation \cite{ekertbook,Nielsen}.
Recently, much attention has been given to relativistic effects in
the context of quantum information theory. Understanding
entanglement in a relativistic setting is important both for
providing a more complete framework for theoretical considerations
and for practical situations such as the implementation of quantum
computation tasks performed by observers in arbitrary relative
motion. So, the relativistic quantum information theory may become
a necessary theory in the near future. Relativistic quantum
information in inertial frames has already been studied
\cite{Terno, Alsing2002,Adami2002,Bergou,li,bartlett}. Peres
\textit{et al} demonstrated that the spin of an electron is not
covariant under Lorentz transformation \cite{Terno}. Alsing and
Milburn \cite{Alsing2002} studied the effect of Lorentz
transformation on maximally spin-entangled Bell states in momentum
eigenstates and Gingrich and Adami \cite{Adami2002} derived a
general transformation rule for the spin-momentum entanglement of
two qubits. Now, it is well known that for different observers in
uniform relative motion the total amount of entanglement is the
same in all inertial frames, although  they don't agree on the
amount of entanglement among various degree of freedom of the
system.

Also, more recently, quantum entanglement has been studied in
relativistic non-inertial frames
\cite{Alsing2003,Mann1,Mann2,Alsing20031,jing,martinez1,martinez2,martinez3,hwang}.
Alsing and Milburn extended the argument to a situation where one
observer is uniformly accelerated \cite{Alsing2003}.
Fuentes-Schuler and Mann calculated the entanglement between two
free modes of a bosonic field, as seen by an inertial observer
detecting one mode and a uniformly accelerated observer detecting
the other mode \cite{Mann1}. Alsing \textit{et al} did this
calculation for two free modes of Dirac field \cite{Mann2}. Pan
and Jing discussed the degradation of entanglement for two
non-maximally entangled free modes of scalar and Dirac fields
\cite{jing}. Mart$\acute{\textrm{i}}$n-Mart$\acute{\textrm{i}}$nez
and Le$\acute{\textrm{o}}$n discussed the behavior of quantum and
classical correlations among all the different spatial-temporal
regions of a spacetime with an event horizon for both fermionic
and bosonic fields \cite{martinez1}. In another work they relaxed
the the single-mode approximation and analyzed bipartite
entanglement degradation due to Unruh effect by introducing an
arbitrary number of accessible modes \cite{martinez2}. Also
Bruschi \textit{et al} addressed the validity of the single-mode
approximation that is commonly invoked in the analysis of quantum
entanglement in noninertial frames. They showed that the
single-mode approximation is not valid for arbitrary states,
finding corrections to previous studies beyond such approximations
in the bosonic and fermionic cases \cite{martinez3}.

In this paper, we attend to tripartite systems which are
constructed with modes of a non-interacting fermionic or bosonic
field. For our purpose, we constraint our argument to the
single-mode approximation
\cite{Alsing2003,Mann1,Mann2,Alsing20031,jing,martinez1}. It is
shown that as a universality principle going beyond this
approximation does not modify the way in which Unruh decoherence
affects fermionic and bosonic entanglements. Unruh effect is
independent of the number of field modes considered in the problem
and statistics is the ruler of this process \cite{martinez3}. In
tripartite discrete systems, two classes of genuine tripartite
entanglement have been discovered, namely, the
Greenberger-Horne-Zielinger (GHZ) class \cite{ghz1,ghz2} and the W
class \cite{w,dur}. These two different types of entanglement are
not equivalent and cannot be converted to each other by local
unitary operations combined with classical communication
\cite{GHZW}. The entanglement in the W state is robust against the
loss of one qubit, while the GHZ state is reduced to a product of
two qubits. According to the geometric measure of entanglement,
the W state has higher entanglement than the GHZ state does
\cite{wei}. Methods are proposed for generation and observation of
GHZ or W type entanglements \cite{sharma,song}.

Here, we consider three observers, two non-inertial observers with
different uniform accelerations and one inertial observer. They
initially share a state built by modes of the fermionic or bosonic
field, as viewed in the Minkowski coordinates. The state is chosen
to be entangled as the GHZ state or the W state. Then switching to
the Rindler coordinates for the accelerated observers, we
investigate the change of entanglement caused by the Unruh effect
\cite{Unruh}. We use the logarithmic negativity as a measure to
evaluate the entanglement. This measure can be used to evaluate
the entanglement of one part relative to the other parts of a
system with arbitrary dimension \cite{Vidal}. In a similar work
Hwang {\it et al} have discussed tripartite entanglement in a
non-inertial frame \cite{hwang}. However, in their problem only
one of the three observers is accelerating and only bosonic
entanglements are discussed.

This paper is organized as follows. In Sec. 2 we review the basic
formalism in brief and introduce the Bogoliubov transformation
both for fermionic and bosonic states. In Sec. 3 we discuss the
Unruh effect for the fermionic GHZ or W entanglements. We use
logarithmic negativity as a measure for the entanglements. The
effect of  Unruh temperature on residual entanglement of bipartite
entanglements is also discussed. In Sec. 4 similar to Sec. 3 we
address the entanglement degradation for Bosonic GHZ or W
entanglements. We demonstrate our results by appropriate 2 or 3
dimensional graphs. Finally in Sec. 5 we present the conclusions.

\section{Basic Formalism}

From an inertial perspective, Minkowski coordinates are the most
suitable coordinates for discussing the problem. However, for a
uniformly accelerated observer the Rindler coordinates are
appropriate. Two different sets of Rindler coordinates are
required for covering the Minkowski space. These two sets, define
two causally disconnected regions. Therefore, the accelerated
observer has access only to one of the regions and so he must
trace over the states in the inaccessible region. This process
leads to an information loss and is known as the Unruh effect.
During the Unruh effect the pure state of one region evolves to a
mixed state and so the entanglement decreases. This effect can be
considered as a local non-unitarian transformation, since
acceleration affect every mod locally. The Unruh effect implies
that a vacuum state in an inertial frame is seen as a thermal
state by an accelerated observer, where the corresponding
temperature depends on the value of acceleration.

Consider a uniformly accelerated observer moving with a proper
acceleration $a$ in Minkowski $(t,z)$ plane. Two different sets of
Rindler coordinates $(\tau,\xi)$ defined as
\begin{eqnarray}\label{Rindlertrans}
      &&a\,t=e^{a\xi}\sinh{a\tau}\,\,\,\,\,
      \qquad a\,z=e^{a\xi}\cosh{a\tau} \,\qquad\text{region
      I}\\\nonumber
     &&a\,t=-e^{a\xi}\sinh{a\tau} \qquad a\,z=-e^{a\xi}\cos{a\tau}
     \qquad\text{region II}
\end{eqnarray}
are necessary for covering Minkowski spacetime \cite{Davies}.
These coordinate transformations define two Rindler regions I and
II that are causally disconnected. In Minkowski coordinates, let
the operators $(\mathcal{A}_k,\mathcal{A}_k^\dag)$ be annihilation
and creation operators for the positive energy solutions and
$(\mathcal{B}_k,\mathcal{B}_k^\dag)$ be annihilation and creation
operators for the negative energy solutions. Here, $k$ is a
notational shorthand for the wave vector $\textbf{k}$, which
labels the modes. Also, in region I, let us denote
$(\mathcal{C}^\textrm{I}_k,\mathcal{C}_k^{\textrm{I}\dag})$ as the
annihilation and creation operators for particles and
$(\mathcal{D}^\textrm{I}_k,\mathcal{D}_k^{\textrm{I}\dag})$ as the
annihilation and creation operators for antiparticles. The
corresponding particle and antiparticle operators in region II are
denoted as
$(\mathcal{C}^{\textrm{II}}_k,\mathcal{C}_k^{\textrm{II}\dag})$
and
$(\mathcal{D}^{\textrm{II}}_k,\mathcal{D}_k^{\textrm{II}\dag})$.

As we know the relationship between the Minkowski and Rindler
creation and annihilation operators is given by the Bogoliubov
transformation. Thus, the Bogoliubov transformation for a
fermionic field is
\begin{equation}
    \left[
    \begin{array}{c}
   \mathcal{B}_k \\
   \mathcal{A}_{-k}^\dagger \\
   \end{array}
   \right] = \left[
   \begin{array}{cc}
   \cos u & \sin u \\
   -\sin u & \cos u \\
   \end{array}
   \right] \left[
   \begin{array}{c}
   \mathcal{D}^{\textrm{I}}_k \\
   \mathcal{C}^{\textrm{II}\dagger}_{-k} \\
   \end{array}
   \right],
\end{equation}
where $\tan u=e^{-\pi \omega c/a}$ with $\omega$ as the particle
frequency and $c$ as the speed of light. Also, for a bosonic field
we have
\begin{equation}
  \left[
  \begin{array}{c}
  \mathcal{B}_k \\
  \mathcal{A}_{-k}^\dagger \\
 \end{array}
 \right] = \left[
 \begin{array}{cc}
  \cosh r & -\sinh r \\
  -\sinh r & \cosh r \\
 \end{array}
 \right] \, \left[
  \begin{array}{c}
  \mathcal{D}^{\textrm{I}}_k \\
  \mathcal{C}^{\textrm{II}\dagger}_{-k} \\
  \end{array}
  \right],
\end{equation}
where $\tanh r=e^{-\pi \omega c/a}$. Note that $0\leq u<\pi/4$ and
$0\leq r<\infty$, as the proper acceleration $a$ takes its full
range, i.e. $0\leq a<\infty$,

Using the above Bogoliubov transformations we can express the
vacuum Minkowski state $|0_k\rangle^+_\mathcal{M}$ and the first
excited state $|1_k\rangle^+_\mathcal{M}$ in terms of Rindler
states. For the fermionic field we can write
\begin{eqnarray}\label{FermionBog}
   |0_k\rangle_\mathcal{M}^+&=&\cos{u}|0_k\rangle_{\textrm{I}}^+|0_{-k}\rangle_{\textrm{II}}^-
   +\sin{u}|1_k\rangle_{\textrm{I}}^+|1_{-k}\rangle_{\textrm{II}}^-\nonumber  \\
   |1_k\rangle_\mathcal{M}^+&=&|1_k\rangle_\textrm{I}^+|0_{-k}\rangle_{\textrm{II}}^-.
\end{eqnarray}
Also, for the bosonic field we have
\begin{eqnarray}\label{BosonBog}
   |0_k\rangle^+_\mathcal{M}&=&\frac{1}{\cosh{r}}\sum_{n=0}^{\infty}\tanh^n{r}|n_k\rangle^+_\textrm{I}|n_{-k}\rangle^-_{\textrm{II}}\nonumber\\
   |1_k\rangle^+_\mathcal{M}&=&\frac{1}{\cosh^2{r}}\sum_{n=0}^{\infty}\tanh^n{r}\sqrt{n+1}|(n+1)_k\rangle^+_\textrm{I}|n_{-k}\rangle^-_{\textrm{II}}.
 \end{eqnarray}
In these transformations $|n_k\rangle^+_\textrm{I}$ and
$|n_{-k}\rangle^-_{\textrm{II}}$ indicate the Rindler-region-I
particle state and Rindler-region-II antiparticle state,
respectively.

\section{Fermionic entanglements}

\subsection{The GHZ state}

\begin{figure}
       \includegraphics[width=6cm,height=5cm]{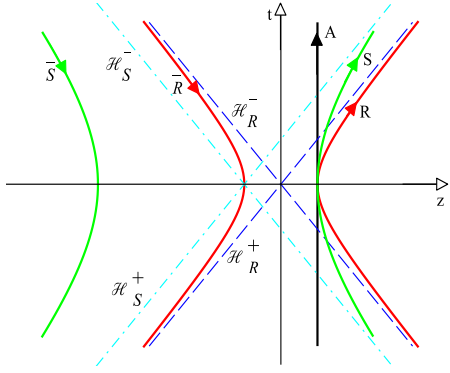}
        \caption{Rindler spacetime diagram for an inertial observer Alice
                  and two accelerated observers Rob and Steven. The thick black line shows the word line
                  of Alice (A). A uniformly accelerated
                  observer Rob (R) with acceleration $a_1$ travels on a
                  hyperbolic world line constrained to region I, and a fictitious observer anti-Rob
                  $(\bar{\textrm{R}})$ travels on the corresponding hyperbola in region II
                  given by the negative of Rob's coordinates (two thick red
                  hyperbola). The horizons of R $(\bar{\textrm{R}})$ are lines of
                  $\tau=\pm\infty$ which Alice (A) will cross them at finite Minkowski
                  times corresponding to $\mathcal{H}^{\pm}_R$ (two dashed blue
                  lines). Another uniformly accelerated observer Steven (S) travels on another hyperbola with
                  an acceleration $a_2$  constrained to region
                  $\textrm{I}'$, and a fictitious observer anti-Steven $(\bar{\textrm{S}})$ travels
                  on the corresponding hyperbola in region $\textrm{II}'$ given by the negative
                  of Steven's coordinates (two thick green hyperbola). The horizons
                  of S $(\bar{\textrm{S}})$, are lines of $\tau'=\pm\infty$ which Alice (A)
                  will cross them at finite Minkowski times, are $\mathcal{H}^{\pm}_S$
                  (two dashed cyan lines).}\label{1FigRindler}
\end{figure}

Consider three observers, Alice, Rob and Steven, such that Alice
is at rest in an inertial frame, but Rob and Steven are moving
with uniform accelerations with respect to Alice. We assign
accelerations $a_1$  and $a_2$ to Steven and Rob, respectively.
Fig. \ref{1FigRindler} shows the corresponding spacetime diagram.
We see that at some points Alice's signals will no longer reach to
Rob and Steven, however, Rob's and Steven's signals will always
reach to Alice. Each observer carries a detector sensitive only to
a single mode of a fermionic field, $k_\textrm{A}$ for Alice,
$k_\textrm{R}$ for Rob and $k_\textrm{S}$ for Steven. We suppose a
GHZ entanglement for this tripartite system, as viewed in an
inertial frame. Therefore, we use the Minkowski modes to construct
a GHZ state for the system as
\begin{equation}\label{GHZket}
      |\textrm{GHZ}\rangle^{\mathcal{M}}_{\textrm{ARS}}=\frac{1}{\sqrt{2}}\left(|0_{k_\textrm{A}}\rangle^+
      |0_{k_\textrm{R}}\rangle^+|0_{k_\textrm{S}}\rangle^+
      +|1_{k_\textrm{A}}\rangle^+|1_{k_\textrm{R}}\rangle^+|1_{k_\textrm{S}}\rangle^+\right).
\end{equation}
where we have used this fact that in a fermionic field there are
only two allowed states for each Minkowski mode.

Now, we need to express the states $|n_{k_\textrm{R}}\rangle^+$
and $|n_{k_\textrm{S}}\rangle^+$, in terms of Rindler states
corresponding to Rob and Steven. Then using (\ref{FermionBog}), we
obtain
\begin{eqnarray}\label{GHZRin}\nonumber
    |\textrm{GHZ}\rangle_{\textrm{ARS}}&=&\frac{1}{\sqrt{2}}\left[|0_{k_\textrm{A}}\rangle^+(\cos
    u_{1}|0_{k_\textrm{R}}\rangle^+_{\textrm{I}}|0_{-k_\textrm{R}}\rangle^- _{\textrm{II}}+
    \sin u_1|1_{k_\textrm{R}}\rangle^+ _{\textrm{I}}|1_{-k_\textrm{R}}\rangle^- _{\textrm{II}})
    (\cos u_{2}|0_{k_\textrm{S}}\rangle^+ _{\textrm{I}'}|0_{-k_S}\rangle^- _{\textrm{II}'}\right. \\&&
    \left.+
    \sin u_{2}|1_{k_\textrm{S}}\rangle^+
    _{I'}|1_{-k_\textrm{S}}\rangle^-_{\textrm{II}'})+|1_{k_\textrm{A}}\rangle^+
    (|1_{k_\textrm{S}}\rangle^+
    _{\textrm{I}}|0_{-k_\textrm{S}}\rangle^- _{\textrm{II}})(|1_{k_\textrm{S}}\rangle^+ _{\textrm{I}'}|0_{-k_\textrm{S}}\rangle^-
    _{\textrm{II}'})\right],
\end{eqnarray}
where $|n_{k_\textrm{R}}\rangle^+_{\textrm{I}}$ (or
$|n_{k_\textrm{S}}\rangle^+ _{\textrm{I}'}$) is the
Rindler-region-I-particle mode for Rob (or Steven),
$|n_{-k_\textrm{R}}\rangle^- _{\textrm{II}}$ (or
$|n_{-k_\textrm{S}}\rangle^- _{\textrm{II}'}$) is
Rindler-region-II-antiparticle mode for Rob (or Steven), and $\tan
u_i=\exp(-\pi\omega c/a_i)$ with $i\in\{1,2\}$.

Using the basis
$\{|n_{k_\textrm{A}}\rangle^+|n_{k_\textrm{R}}\rangle^+|n_{k_\textrm{S}}\rangle^+\}$
that is
$\{|000\rangle,|001\rangle,|010\rangle,|100\rangle,|011\rangle,|101\rangle,|110\rangle,|111\rangle\}$,
we can obtain the density matrix for the GHZ state (\ref{GHZket})
as
\begin{eqnarray}\label{GHZMin}
      \rho^{\mathcal{M}}_{\textrm{ARS}}&=&\frac{1}{8}
      \left(
    \SSS{1}{1}{1}-\SSS{1}{2}{2}-\SSS{2}{1}{2}-\SSS{2}{2}{1}\right.\nonumber
    \\ &&\left.+\Sigma_3\otimes\Sigma_3\otimes I+\Sigma_3\otimes I\otimes
    \Sigma_3+I\otimes\Sigma_3\otimes \Sigma_3+I\otimes I \otimes I
      \right)
\end{eqnarray}
where $I$ is the unit $2\times 2$ matrix and
$\{\Sigma_1,\Sigma_2,\Sigma_3\}$ are the Pauli matrices. Notice
that here and in the following, the matrices in each tensor
product term are placed according to Alice-Rob-Steven order. The
pure state (\ref{GHZMin}) describes a tripartite system. On the
other hand, the density matrix for the state (\ref{GHZRin}) is
obtained as
$\rho_{\textrm{A},\textrm{I},\textrm{I}',\textrm{II},\textrm{II}'}=|\textrm{GHZ}\rangle_{\textrm{ARS}}\langle
\textrm{GHZ}|$, which is pure and describes a five-partite system.
However, as Fig. 1 represents, whole of spacetime is accessible
only for the inertial observer Alice, and the accelerating
observer Rob (or Steven) has only access to one region say I (or
$\textrm{I}'$). So, we must trace over the states belong to
regions II and $\textrm{II}'$. Doing so, we get the following
tripartite density matrix
\begin{eqnarray}\label{GHZRin}\nonumber
     \rho_{\textrm{A},\textrm{I},\textrm{I}'}&=&\frac{1}{8}\left[\cos{u_1}\cos{u_2}(\SSS{1}{1}{1}-\SSS{1}{2}{2}-\SSS{2}{1}{2}\right.\\\nonumber
    &&\left.-\SSS{2}{2}{1})+
    \frac{1}{2}(\cos{2u_1}\cos{2u_2}-1)(\SSS{3}{3}{3})+\cos^2{u_1}(\Sigma_3\otimes\Sigma_3\otimes
    I)\right.\\\nonumber &&\left.+\cos^2 u_2(\Sigma_3\otimes
    I\otimes \Sigma_3)
    +\frac{1}{2}(\cos{2u_1}\cos{2u_2}+1)(I\otimes\CC{3}{3})-\sin^2 u_1(I\otimes\Sigma_3\otimes
    I)\right.\\
    &&\left.-\sin^2u_2(I\otimes
    I \otimes\Sigma_3)+I\otimes
    I \otimes
    I\right]
\end{eqnarray}
where we have employed the basis
$\{|n_{k_\textrm{A}}\rangle^+|n_{k_\textrm{R}}\rangle^+_{\textrm{I}}|n_{k_\textrm{S}}\rangle^+
_{\textrm{I}'}\}$. While the state (\ref{GHZMin}) is pure and
maximally entangled, the state (\ref{GHZRin}) is not pure and, as
we will verify, its entanglement is degraded. This entanglement
degradation is essentially justified by the Unruh effect.

\subsubsection{A-RS, R-AS, S-AR entanglements}

To quantify the entanglement in a tripartite system, different
measures have been introduced. Here, regarding the dimension of
the density matrix (\ref{GHZRin}) and that the state is not pure,
we consider the logarithmic negativity which describes the
entanglement of one part of the system relative to the other
parts. For instance, the logarithmic negativity of Alice part
relative to the other two parts is defined as
$\mathcal{N}_{\textrm{A}-\textrm{RS}}=\sum_i\log_2|\lambda_i|$
where $\lambda_i$ denotes the  eigenvalues of
$\rho_{\widetilde{\textrm{A}},\textrm{I},\textrm{I}'}$ which is
the partial transposition of
$\rho_{\textrm{A},\textrm{I},\textrm{I}'}$ with respect to Alice.
Similarly, we can calculate $\mathcal{N}_{\textrm{R}-\textrm{AS}}$
and $\mathcal{N}_{\textrm{S}-\textrm{AR}}$ by finding the
eigenvalues of
$\rho_{\textrm{A},\widetilde{\textrm{I}},\textrm{I}'}$ (partially
transposed density matrix with respect to Rob) and
$\rho_{\textrm{A},\textrm{I},\widetilde{\textrm{I}'}}$ (partially
transposed density matrix with respect to Steven). Logarithmic
Negativity vanishes unless some negative eigenvalues are present.
Let $N$ denote the negative eigenvalue, then we can write the
logarithmic negativity also as
\begin{equation}\label{LogNeg}
      \mathcal{N}=\log_2(1-2N)
\end{equation}

\begin{figure}
\subfigure[ ]{
    \label{LN3GHZ:subfig:a}     \begin{minipage}[b]{0.4\textwidth}
      \centering
      \includegraphics[width=6cm,height=6cm]{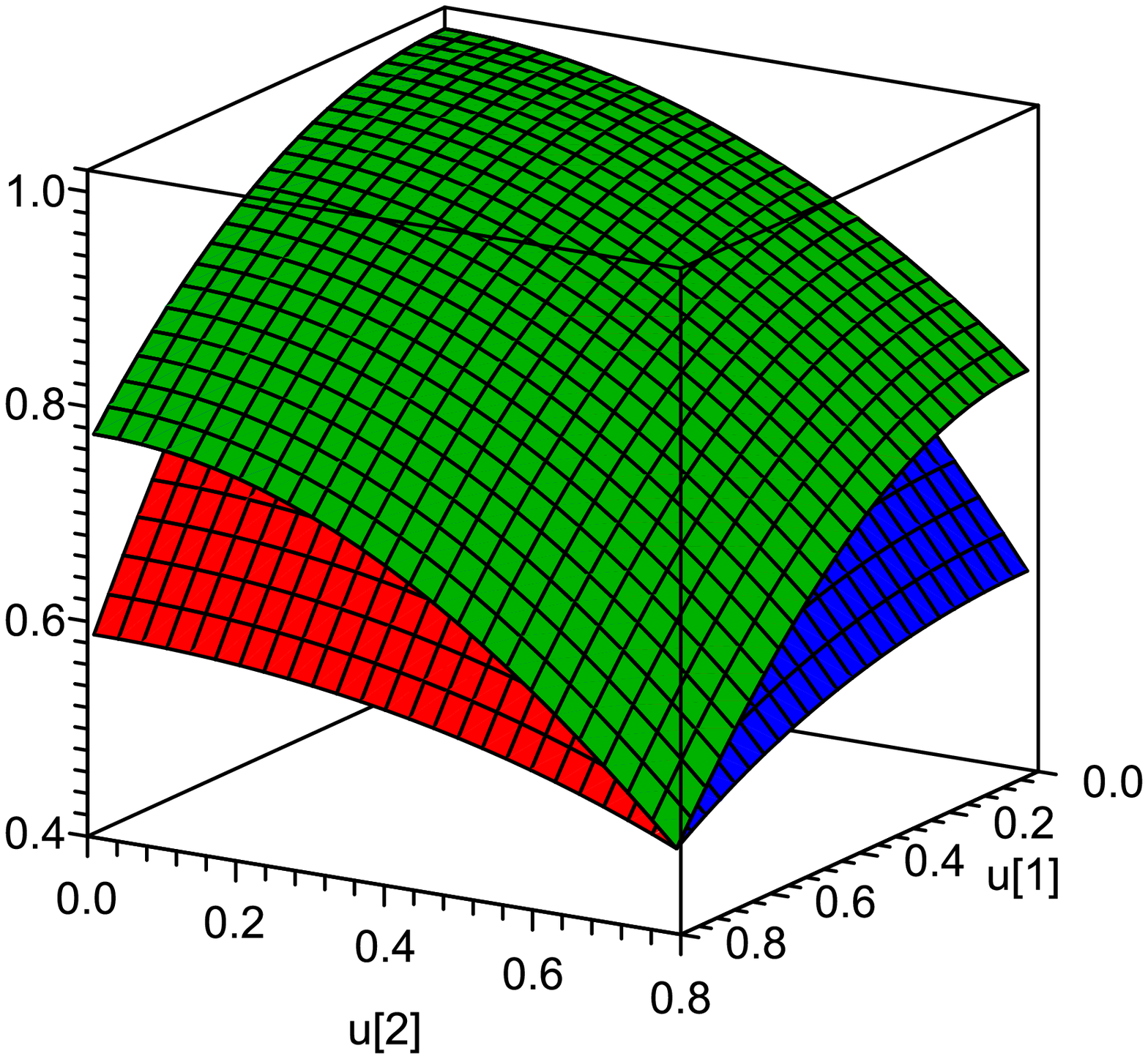}
    \end{minipage}}
\subfigure[]{
    \label{LN3GHZ:subfig:b}     \begin{minipage}[b]{0.4\textwidth}
      \centering
      \includegraphics[width=6cm,height=6cm]{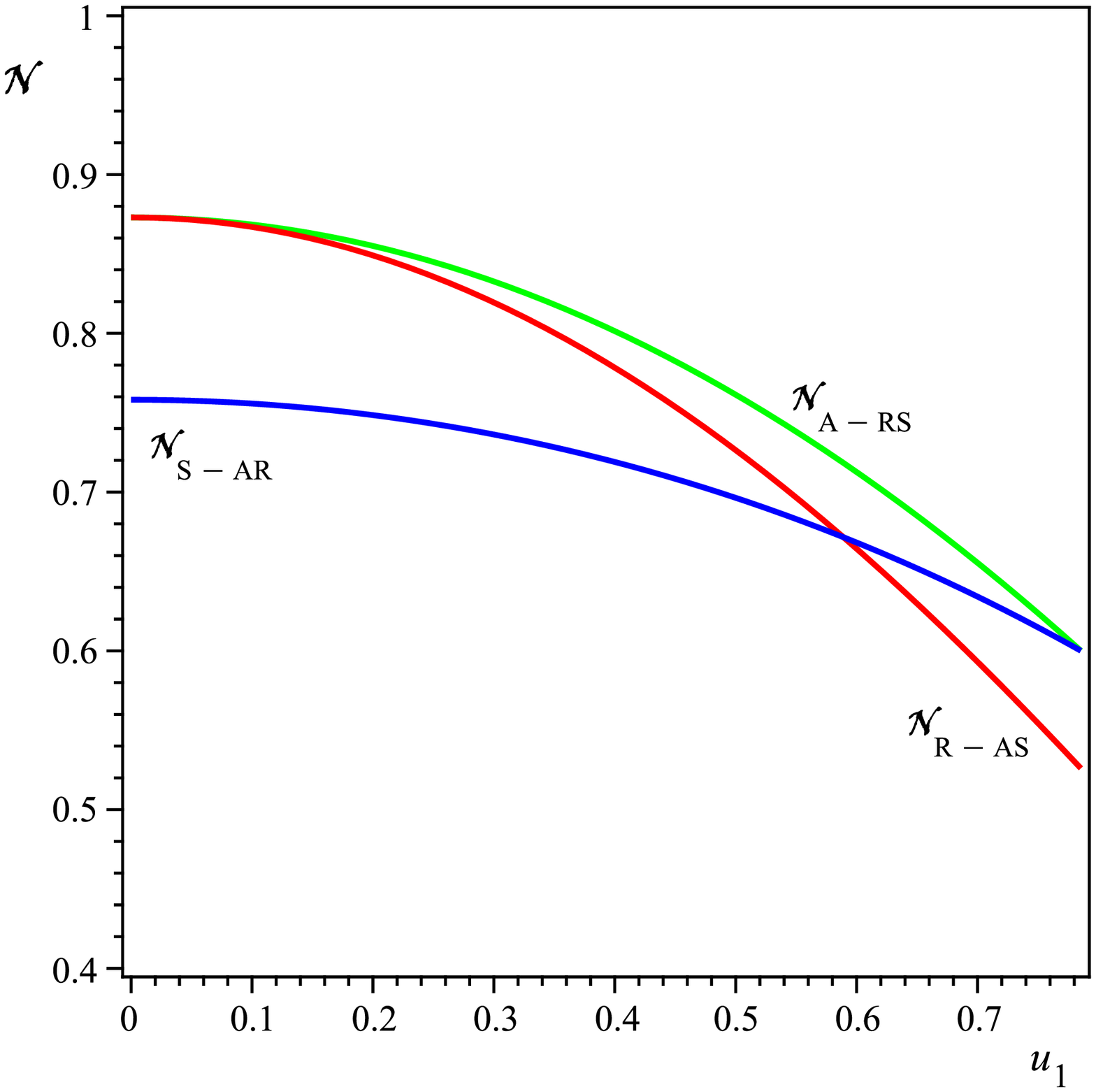}
    \end{minipage}}
\caption{Logarithmic negativity for the fermionic GHZ
  state versus accelerations $u_1$ and $u_2$. (a) The two lower surfaces that intersect at $u_1=u_2$,
represent $\mathcal{N}_{\textrm{R}-\textrm{AS}}$ and
$\mathcal{N}_{\textrm{S}-\textrm{AR}}$. The upper surface
corresponds to $\mathcal{N}_{\textrm{A}-\textrm{RS}}$. (b) A
section of figure (a) for a given $u_2=3\pi/16$. Even for infinite
accelerations or $u_1, u_2\rightarrow\frac{\pi}{4}$, each
logarithmic negativity does not vanish, i.e., the Unruh effect
does not completely destroy the entanglement in this case.}
\label{LN3GHZ}
\end{figure}

We can readily obtain the required partially transposed matrices
from (\ref{GHZRin}) by noting that after a transposition,
$\Sigma_1$ and $\Sigma_3$ do not change but
$\Sigma_2\longrightarrow -\Sigma_2$. We have
\begin{equation}\label{FGHZPA}
       \rho_{\widetilde{\textrm{A}},\textrm{I},\textrm{I}'}=\varrho-\frac{1}{8}\cos{u_1}\cos{u_2}
       (\SSS{1}{2}{2}-\SSS{2}{1}{2}-\SSS{2}{2}{1}),
\end{equation}
\begin{equation}\label{FGHZPR}
      \rho_{\textrm{A},\widetilde{\textrm{I}},\textrm{I}'}=\varrho-\frac{1}{8}\cos{u_1}\cos{u_2}
       (-\SSS{1}{2}{2}+\SSS{2}{1}{2}-\SSS{2}{2}{1}),
\end{equation}
and
\begin{equation}\label{FGHZPS}
      \rho_{\textrm{A},\textrm{I},\widetilde{\textrm{I}'}}=\varrho-\frac{1}{8}\cos{u_1}\cos{u_2}
       (-\SSS{1}{2}{2}-\SSS{2}{1}{2}+\SSS{2}{2}{1}),
\end{equation}
where
\begin{eqnarray}
      \varrho&=&\frac{1}{8}\left[\cos{u_1}\cos{u_2}(\SSS{1}{1}{1})+
    \frac{1}{2}(\cos{2u_1}\cos{2u_2}-1)(\SSS{3}{3}{3})\right.\\ \nonumber&&\left.+\cos^2{u_1}(\Sigma_3\otimes\Sigma_3\otimes
    I)+\cos^2 u_2(\Sigma_3\otimes
    I\otimes \Sigma_3)
    +\frac{1}{2}(\cos{2u_1}\cos{2u_2}+1)(I\otimes\CC{3}{3})\right.\\\nonumber
    &&\left.-\sin^2 u_1(I\otimes\Sigma_3\otimes
    I)-\sin^2u_2(I\otimes
    I \otimes\Sigma_3)+I\otimes
    I \otimes
    I\right].
\end{eqnarray}
The eigenvalues of each of these partially transposed matrices can
be obtained explicitly, however, we need only negative
eigenvalues. It turns out that the negative eigenvalues for
(\ref{FGHZPA}), (\ref{FGHZPR}) and (\ref{FGHZPS}) are
\begin{equation}
      N_\textrm{A-RS}=\frac{1}{4}\sin^{2} u_1 \sin^{2}
      u_2-\frac{1}{4}\,\sqrt{\sin^{4} u_1
      \sin^{4}u_2+4\cos^2u_1\cos^2u_2},
\end{equation}
\begin{equation}\label{NRAS}
      N_\textrm{R-AS}=\frac{1}{4}\sin^{2} u_1 \cos^{2}
      u_2-\frac{1}{4}\,\sqrt{\sin^{4} u_1
      \cos^{4}u_2+4\cos^2u_1\cos^2u_2},
\end{equation}
and
\begin{equation}\label{NRAS}
      N_\textrm{S-AR}=\frac{1}{4}\sin^{2} u_2 \cos^{2}
      u_1-\frac{1}{4}\,\sqrt{\sin^{4} u_2
      \cos^{4}u_1+4\cos^2u_2\cos^2u_1},
\end{equation}
respectively. For investigating the entanglement, we substitute
these eigenvalues in (\ref{LogNeg}) to obtain the corresponding
logarithmic negativity as functions of $u_1$ and $u_2$. Recall
that in the present case $0\leq u_i<\frac{\pi}{4}$ corresponding
to $0\leq a_i<\infty$ where $a_i$ indicates the proper
acceleration of Rob or Steven. We have plotted all three surfaces
$\mathcal{N}_{\textrm{R}-\textrm{AS}}$,
$\mathcal{N}_{\textrm{S}-\textrm{AR}}$ and
$\mathcal{N}_{\textrm{A}-\textrm{RS}}$ in Fig.
\ref{LN3GHZ:subfig:a}. Each surface shows a uniform decreasing of
logarithmic negativity as accelerations increase. It must be noted
that even for infinite accelerations, that is, $u_1,
u_2\rightarrow\frac{\pi}{4}$, each logarithmic negativity does not
vanish. In other words, like bipartite entanglement, the Unruh
effect doesn't completely destroy the entanglement. We see that
the surface of $\mathcal{N}_{\textrm{A}-\textrm{RS}}$ covers the
other two surfaces. To illustrate the situation a section of Fig.
\ref{LN3GHZ:subfig:a} for a given $u_2=3\pi/16$ is plotted in Fig.
\ref{LN3GHZ:subfig:b}. It is remarkable that these entanglements
are not generally equal. This means that in the considered
tripartite system each part is differently entangled to the other
two parts; the inertial Alice part is mostly entangled. Of course,
the surfaces of $\mathcal{N}_{\textrm{R}-\textrm{AS}}$ and
$\mathcal{N}_{\textrm{S}-\textrm{AR}}$ intersect at $u_1=u_2$,
i.e., when Rob and Steven have the same acceleration, have the
same entanglement, expectedly.

\subsubsection{Entanglement of bipartite subsystems}

In this subsection, we attend to the entanglement of bipartite
subsystems AR, AS and RS. To do this, we must trace out one of the
parts of the tripartite system ARS described by the density matrix
(\ref{GHZRin}). We can readily trace over Alice, Rob or Steven
states by noting that the Pauli matrices are traceless. For
instance, upon tracing over the Alice states, only the tensor
product terms remain where the first matrix is the unity matrix
$I$. Then, we obtain the following reduced density matrix for RS
subsystem
\begin{equation}\label{DDGHZ}
      \rho_{\textrm{I},\textrm{I}'} =\frac{1}{4}\left[\frac{1}{2}(\cos{2u_1}\cos{2u_2}+1)
      (\Sigma_3\otimes\Sigma_3)-\sin^2u_1(\Sigma_3\otimes I)-\sin^2u_2(I\otimes\Sigma_3)+I\otimes I\right],
\end{equation}
which is diagonal. In the same manner we see that the reduced
density matrices for AR and AS subsystems are also diagonal. Thus,
each of these bipartite subsystems is disentangled. This means
that the GHZ character of the state is preserved under the Unruh
effect, that is, tracing over any part of a GHZ state, leads to a
disentangled bipartite subsystem \cite{GHZW}.

\subsection{The  W state}

In this subsection we assume a W entanglement for the tripartite
system ARS as viewed in an inertial frame. Therefor, we use the
Minkowski modes to construct a W state as
\begin{widetext}
\begin{equation}\label{Wket}
      |\textrm{W}\rangle^{\mathcal{M}}_{\textrm{ARS}} =
      \frac{1}{\sqrt{3}}\left(|1_{k_\textrm{A}}\rangle^+|0_{k_\textrm{R}}\rangle^+|0_{k_\textrm{S}}\rangle^+
      +|0_{k_\textrm{A}}\rangle^+|1_{k_\textrm{R}}\rangle^+|0_{k_\textrm{S}}\rangle^+
      +|0_{k_\textrm{A}}\rangle^+|0_{k_\textrm{R}}\rangle^+|1_{k_\textrm{S}}\rangle^+\right).
\end{equation}
\end{widetext}
The corresponding density matrix for this state is obtained as
\begin{eqnarray}\label{WMin}
      \rho^{\mathcal{M}}_{\textrm{ARS}}&=&\frac{1}{24}
       \left[2(\SSS{1}{1}{3}+\SSS{2}{2}{3}+\SSS{1}{3}{1}+\SSS{2}{3}{2}\right.\\\no
       &&\left.+\SSS{3}{1}{1}+\SSS{3}{2}{2}+ \SIS{1}{1}+\SSI{1}{1}+\SSI{2}{2}\right.\\\no
        &&\left.+\SIS{2}{2}+\ISS{1}{1} +\ISS{2}{2})-3(\SSS{3}{3}{3})-\SIS{3}{3}\right.\\\no
        &&\left.-\ISS{3}{3}-\SSI{3}{3}+\Sigma_{3}\otimes I\otimes I+I\otimes\Sigma_{3}\otimes I +I\otimes I \otimes\Sigma_3
       +3(I\otimes I\otimes I)\right].
\end{eqnarray}
Then we apply the Bogoliubov transformation (\ref{FermionBog}) for
the accelerating observers Rob and Steven and rewrite the state
(\ref{Wket}) in Rindler coordinates which leads to a density
matrix describing a five-partite system. After tracing over the
causally disconnected regions II and $\textrm{II}'$, we reach to
\begin{eqnarray}\label{WRind}\no
      \rho_{\textrm{A},\textrm{I},\textrm{I}'}&=&
      \frac{1}{24}\left[2\cos{u_1}\cos{2u_2}(\SSS{1}{1}{3}+\SSS{2}{2}{3})+2\cos{u_2}(\SIS{1}{1}+\SIS{2}{2})\right.\\\no
      &&\left.+2\cos{u_1}(\SSI{1}{1}+\SSI{2}{2})+2\cos{2u_1}\cos{u_2}(\SSS{1}{3}{1}+\SSS{2}{3}{2})\right.\\\no
      &&\left.+2\cos{u_1}\cos{u_2}(\SSS{3}{1}{1}+\SSS{3}{2}{2}+\ISS{1}{1}+\ISS{2}{2})\right.\\\no
      &&\left.-(\cos2u_1\cos2u_2+\cos2{u_1}+\cos2{u_2})(\SSS{3}{3}{3})-\SSI{3}{3}-\SIS{3}{3}+\Sigma_{3}\otimes I \otimes I\right.\\\no
      &&\left.+(\cos 2u_1 \cos2u_2-\cos2u_1-\cos2u_2)(\ISS{3}{3})+(2\cos2{u_1}-1)(I\otimes\Sigma_{3}\otimes I)\right.\\
      &&\left.+(2\cos2{u_2}-1)(I\otimes I \otimes \Sigma_3)+3(I\otimes I\otimes
      I)\right]
\end{eqnarray}

\subsubsection{A-RS, R-AS and  S-AR entanglements}

\begin{figure}
  \subfigure[]{
    \label{FWPA:subfig:a} 
    \begin{minipage}[h]{0.4\textwidth}
      \centering
      \includegraphics[width=5cm,height=5cm]{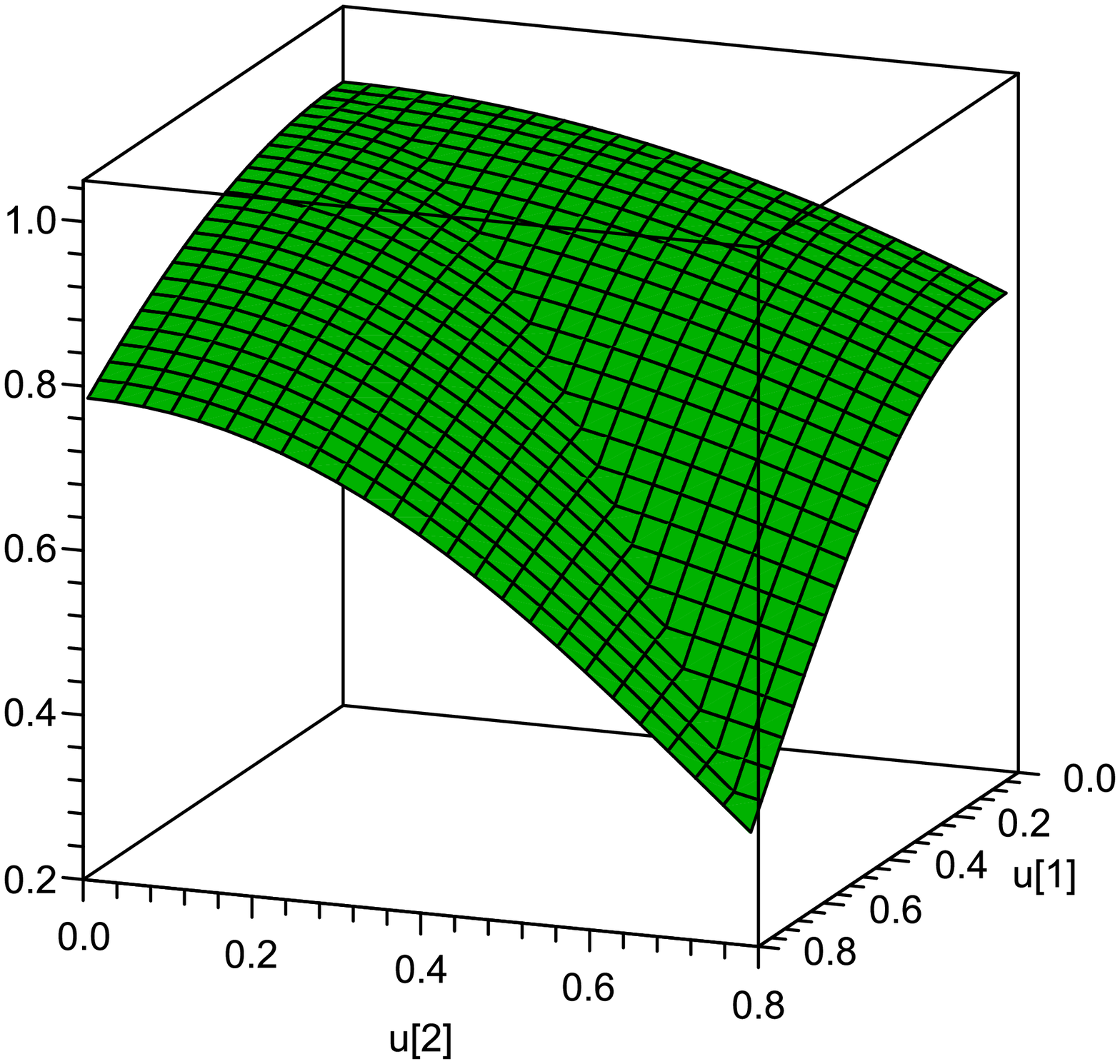}
    \end{minipage}}
     \subfigure[]{
    \label{FW3:subfig:b}
    \begin{minipage}[h]{0.4\textwidth}
      \centering
      \includegraphics[width=6cm,height=6cm]{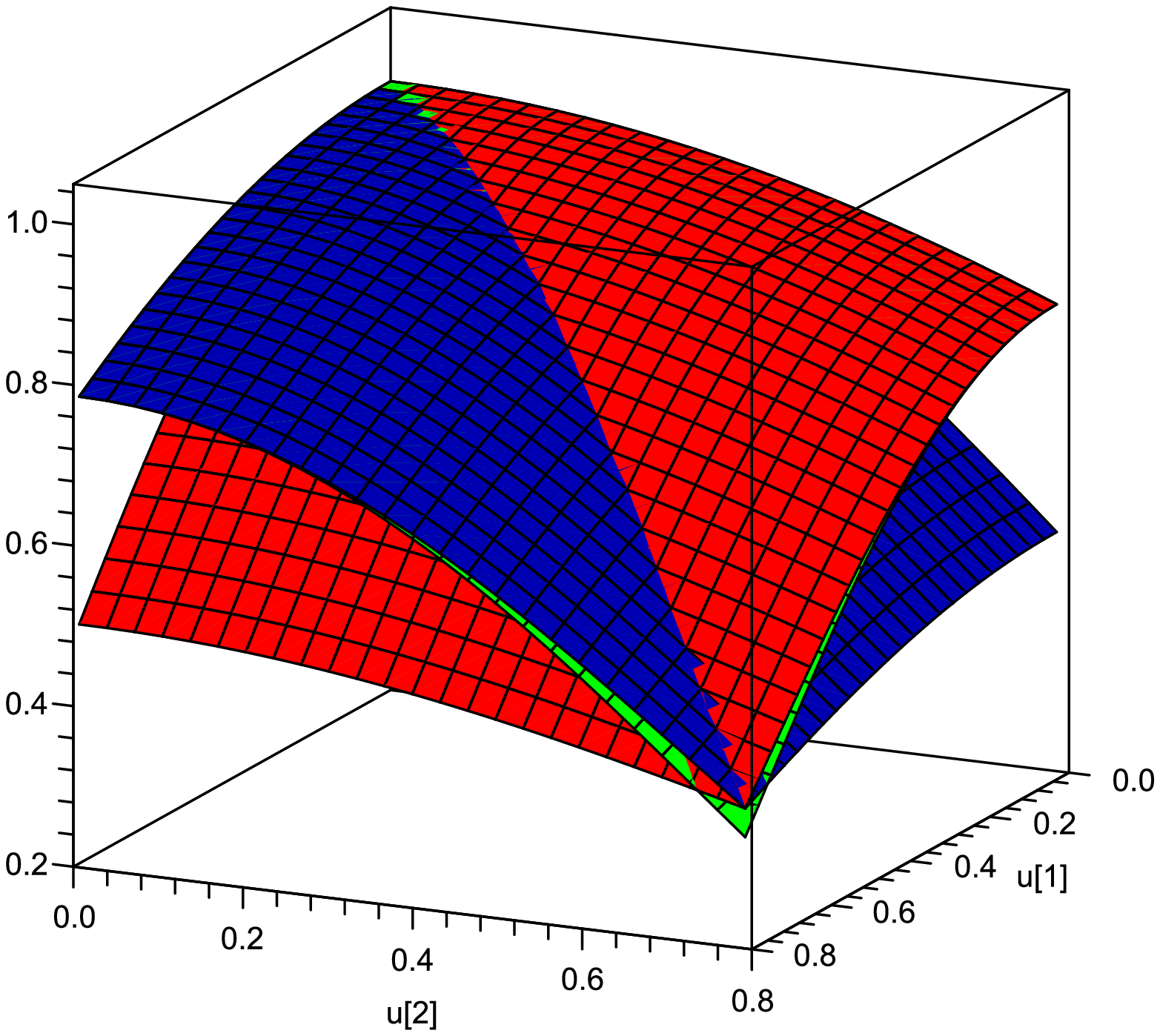}
    \end{minipage}}
    \subfigure[]{
    \label{FWPA:subfig:c} 
    \begin{minipage}[h]{0.4\textwidth}
      \centering
      \includegraphics[width=5cm,height=5cm]{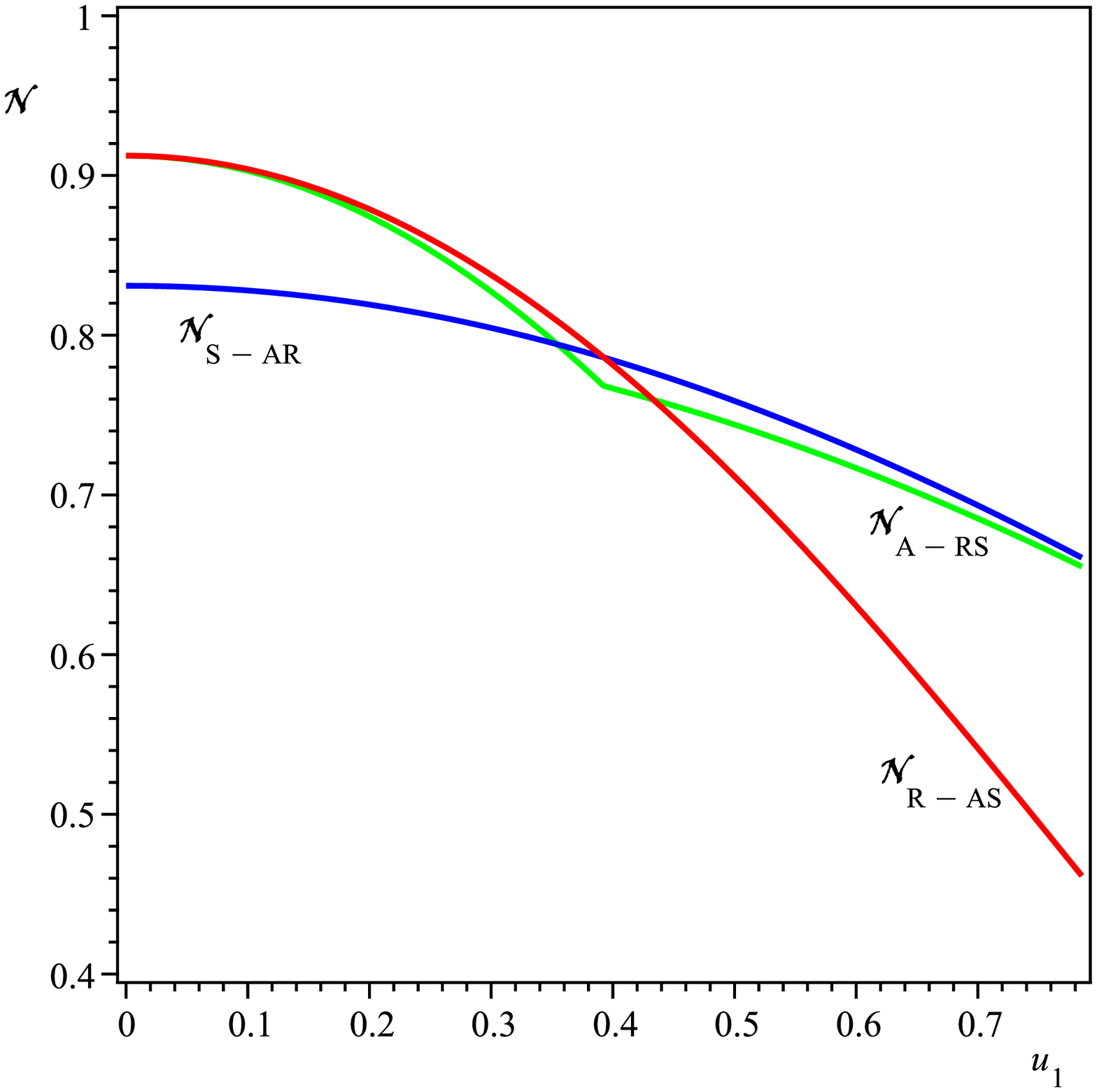}
    \end{minipage}}
  \caption{Logarithmic negativity for the fermionic W
  state versus accelerations $u_1$ and $u_2$. (a) The surface
   $\mathcal{N}_{\textrm{A}-\textrm{RS}}$ in terms of $u_1$ and $u_2$.
   (b) $\mathcal{N}_{\textrm{A}-\textrm{RS}}$, $\mathcal{N}_{\textrm{R}-\textrm{AS}}$
   and $\mathcal{N}_{\textrm{S}-\textrm{AR}}$ surfaces all together. Again
  $\mathcal{N}_{\textrm{R}-\textrm{AS}}$ and $\mathcal{N}_{\textrm{S}-\textrm{AR}}$ surfaces intersect along
  $u_1=u_2$. The surface of $\mathcal{N}_{\textrm{A}-\textrm{RS}}$ is not seen here
  , because, this lies below the other two surfaces. (c) A section of the figure (b)
  for a given $u_2=\pi/8$. The surface $\mathcal{N}_{\textrm{A}-\textrm{RS}}$ intersects
  $\mathcal{N}_{\textrm{R}-\textrm{AS}}$ or $\mathcal{N}_{\textrm{S}-\textrm{AR}}$
  at determined values of $u_1$ and $u_2$. Like the GHZ case, even for infinite
values of $a_1$ and $a_2$, the logarithmic negativity generally is
nonzero. }
  \label{FWLN}
\end{figure}

Again we want to use the logarithmic negativity given by
(\ref{LogNeg}) for evaluating the entanglement of the state
(\ref{WRind}). To do this we first obtain the partially transposed
matrices
$\rho_{\widetilde{\textrm{A}},\textrm{I},\textrm{I}'}$,$\rho_{\textrm{A},\widetilde{\textrm{I}},\textrm{I}'}$
and $\rho_{\textrm{A},\textrm{I},\widetilde{\textrm{I}'}}$. Then
we must calculate the negative eigenvalues for them and substitute
in (\ref{LogNeg}). However, these calculations are lengthy and
here we only present the results by plotting the obtained
logarithmic negativity in Fig. \ref{FWLN}. These surfaces show the
logarithmic negativity in term of accelerations $u_1$ and $u_2$.
We see that the surface of $\mathcal{N}_{\textrm{A}-\textrm{RS}}$
lies below of the surfaces of
$\mathcal{N}_{\textrm{R}-\textrm{AS}}$ and
$\mathcal{N}_{\textrm{S}-\textrm{AR}}$, and is covered by them
when $u_1$ and $u_2$ take their full range. Fig.
(\ref{FWPA:subfig:c}) represent the situation more clearly. As is
seen, again in the considered system each part is entangled
differently. Expectedly, the surfaces
$\mathcal{N}_{\textrm{R}-\textrm{AS}}$ and
$\mathcal{N}_{\textrm{S}-\textrm{AR}}$ intersect at $u_1=u_2$.
However, it is interesting to note that the surface
$\mathcal{N}_{\textrm{A}-\textrm{RS}}$ intersects the
$\mathcal{N}_{\textrm{R}-\textrm{AS}}$ and
$\mathcal{N}_{\textrm{S}-\textrm{AR}}$ at definite values of $u_1$
and $u_2$, i.e., in the considered W state, for definite
accelerations the entanglement of part R or S  can be equal to the
entanglement of  inertial part A. Note that similar to the GHZ
case discussed in the previous subsection, some degree of W
entanglement is preserved as  $a_1$ and $a_2$ go to infinity. This
result is obtained also for fermionic bipartite entanglement and
has been proven to be universal for fermionic fields \cite{
martinez2}.

\subsubsection{Entanglement of bipartite subsystems}

\begin{figure}
  \subfigure[ ]{
    \label{LN3W2a}
    \begin{minipage}[b]{0.4\textwidth}
      \centering
      \includegraphics[width=6cm,height=6cm]{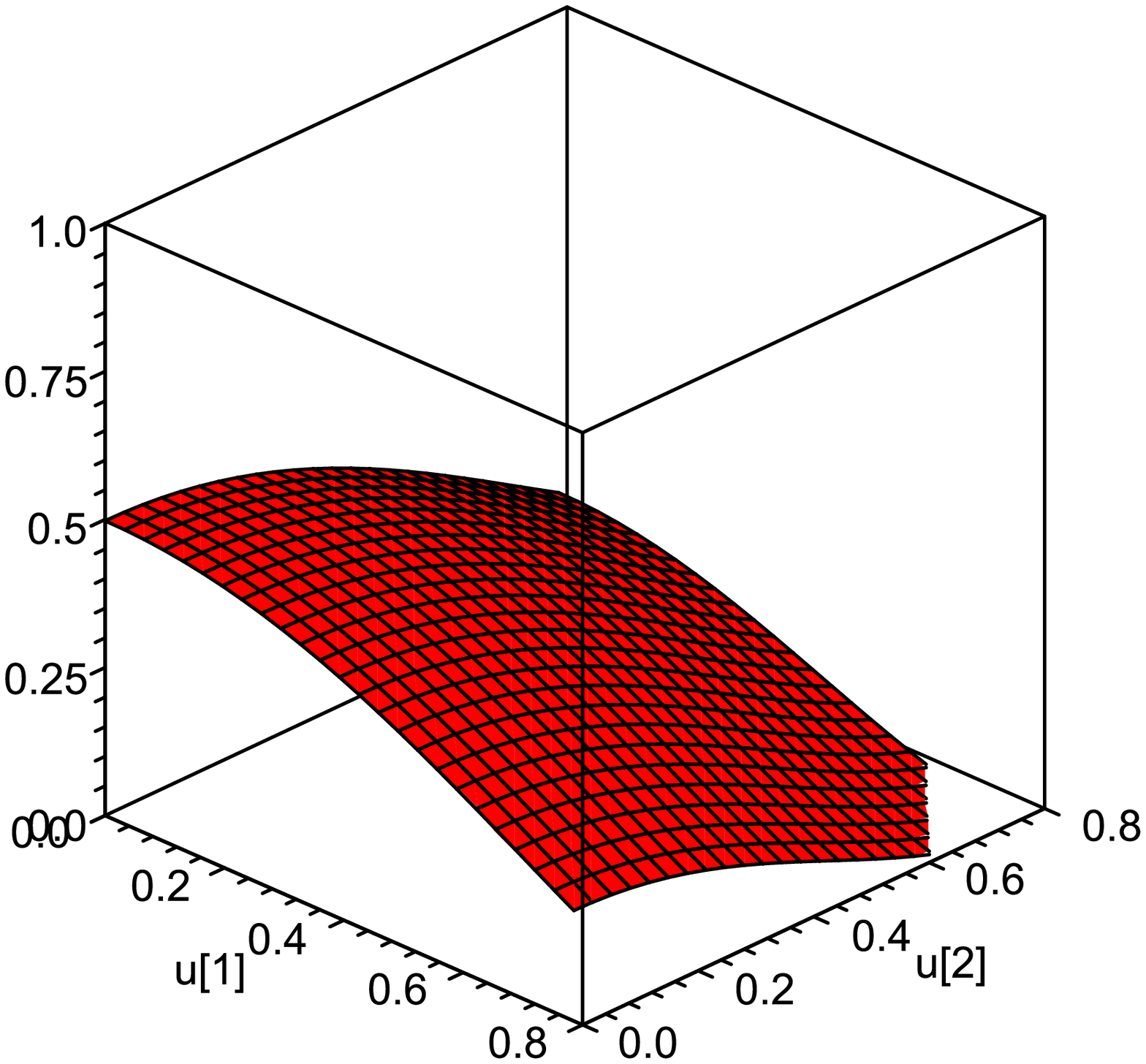}
    \end{minipage}}
  \subfigure[]{
    \label{LN3W2b}
    \begin{minipage}[b]{0.4\textwidth}
      \centering
      \includegraphics[width=6cm,height=6cm]{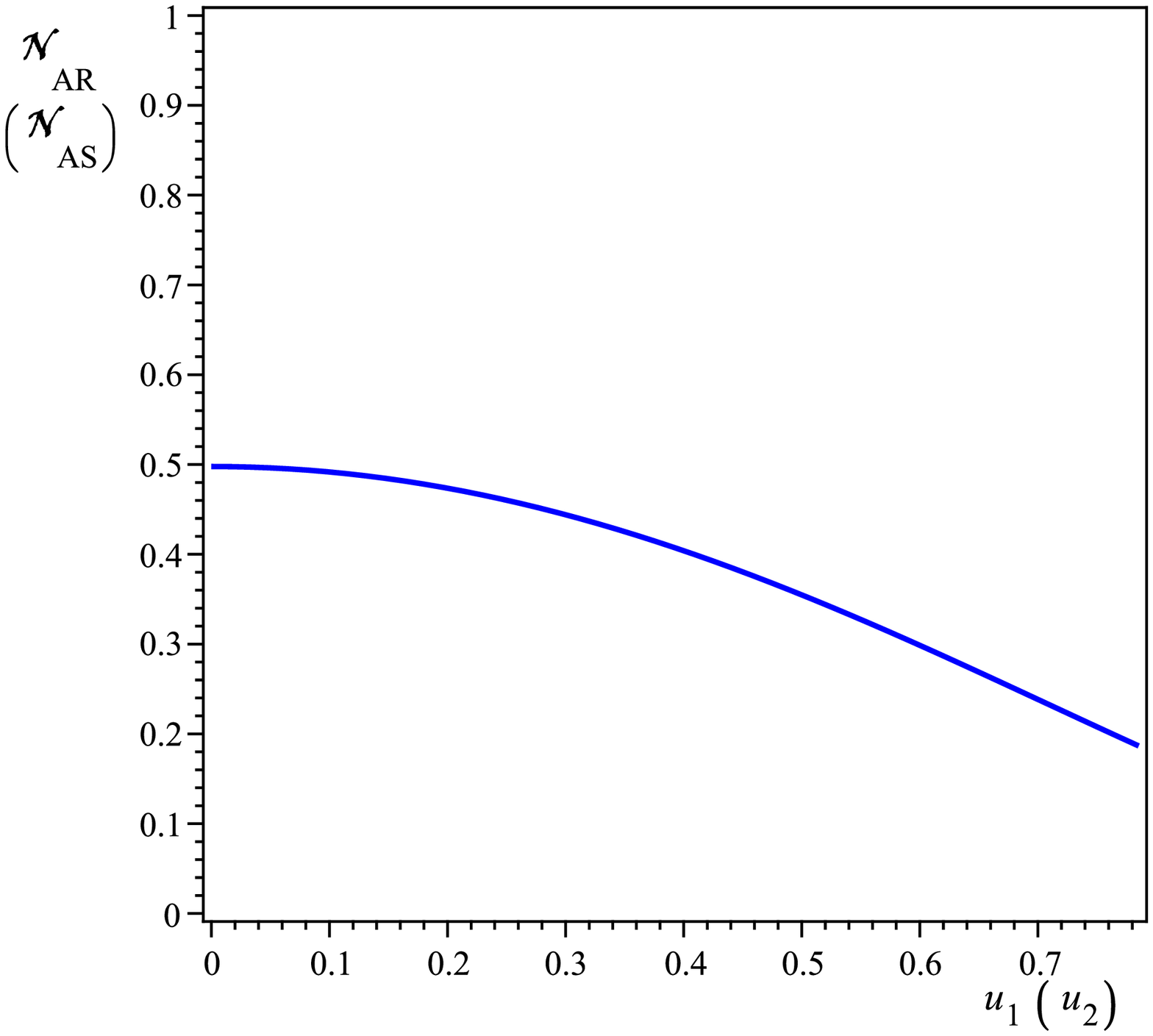}
    \end{minipage}}
     \caption{The logarithmic negativity versus $u_1$ and $u_2$ for bipartite subsystems when the tripartite system
     is in the fermionic W state.
     (a) $\mathcal{N}_{\textrm{R}\textrm{S}}$ surface which represents the entanglement of Rob and Steven states in their
     first regions $\textrm{I}$ and $\textrm{I}'$. As is seen, $\mathcal{N}_{\textrm{R}\textrm{S}}$
     equals $   0.5$ at $u_1=u_2=0$, then reaches to zero at a curve defined
     by (\ref{root}). (b) $\mathcal{N}_{\textrm{A}\textrm{R}}$
     ($\mathcal{N}_{\textrm{A}\textrm{S}}$) depends only on $u_1$ ($u_2$). The curve starts with $   0.5$
     at $u_1=0$ ($u_2=0$), then descends to a non-zero asymptotic value. }
  \label{LN3W2}
\end{figure}

Let us study the entanglement of bipartite subsystems AR, AS and
RS, by tracing over the states of one of the parts of the
tripartite system described by (\ref{WRind}).  We denote the
resulting density matrices as $\rho_{\textrm{I},\textrm{I}'}$,
$\rho_{\textrm{A},\textrm{I}}$ and
$\rho_{\textrm{A},\textrm{I}'}$, when we trace over Alice, Steven
and Rob, respectively. We readily obtain from (\ref{WRind})
\begin{eqnarray}\label{RS}\no
         \rho_{\textrm{I},\textrm{I}'}&=&\frac{1}{12}\left[
        2\cos{u_1}\cos{u_2}(\CC{1}{1}+\CC{2}{2})
        +(\cos 2u_1 \cos2u_2-\cos2u_1-\cos2u_2)(\CC{3}{3})\right.\\
         &&\left.+(2\cos2u_1-1)(\Sigma_3\otimes I)+(2\cos2u_2-1)(I\otimes \Sigma_3)+3(I\otimes I)\right],
\end{eqnarray}
\begin{eqnarray}\label{AR}
        \rho_{\textrm{A},\textrm{I}}&=&\frac{1}{12}\left[ 2\cos{u_1}(\CC{1}{1}+\CC{2}{2})-\CC{3}{3}+\Sigma_3\otimes
       I\right.\\\no
         &&+\left.(2\cos2u_1-1)(I\otimes\Sigma_{3})+3(I\otimes I) \right],
\end{eqnarray}
and
\begin{eqnarray}\label{AS}
        \rho_{\textrm{A},\textrm{I}'}&=&\frac{1}{12}\left[ 2\cos{u_2}(\CC{1}{1}+\CC{2}{2})-\CC{3}{3}+\Sigma_3\otimes
       I\right.\\\no
         &&+\left.(2\cos2u_2-1)(I\otimes\Sigma_{3})+3(I\otimes
        I) \right],
\end{eqnarray}
In contrast to the GHZ case, these bipartite subsystems are not
disentangled. We can still use the logarithmic negativity
(\ref{LogNeg}) for calculating these bipartite entanglements. To
do this we must first find the negative eigenvalues for partially
transposed matrices $\rho_{\widetilde{\textrm{I}},\textrm{I}'}$,
$\rho_{\widetilde{\textrm{A}},\textrm{I}}$ and
$\rho_{\widetilde{\textrm{A}},\textrm{I}}$ corresponding to the
above matrices. We obtain
\begin{eqnarray}\label{nrs}\nonumber
        N_{\textrm{R}\textrm{S}}&=&\frac{1}{2}-\frac{1}{3}(\cos^2u_1+\cos^2u_2)+\frac{1}{3}( \cos u_1 \cos u_2 )^2\\
        &&-\frac{1}{6}\sqrt {9-12(\cos^2 u_1+\cos^2 u_2)+12(\cos u_1\cos u_2) ^{2}+4(\cos^4
        u_1+\cos^4 u_2)},
\end{eqnarray}
\begin{equation}
      N_{\textrm{A}\textrm{R}}=\frac{1}{6}-\frac{1}{6}\,\sqrt{1+4\,\cos^4u_2},
\end{equation}
\begin{equation}
      N_{\textrm{A}\textrm{S}}=\frac{1}{6}-\frac{1}{6}\,\sqrt{1+4\,\cos^4u_1},
\end{equation}
which as substituted in (\ref{LogNeg}), give the corresponding
logarithmic negativity. It is remarkable that the negativity
$N_{\textrm{R}\textrm{S}}$ in (\ref{nrs}) and consequently the
logarithmic negativity $\mathcal{N}_{\textrm{R}\textrm{S}}$ vanish
if $u_1$ and $u_2$ satisfy the equation
\begin{equation}\label{root}
     \cos u_2=\frac{\sqrt{2}\sin u_1}{\sqrt{2-\cos^2u_1}}.
\end{equation}
Fig. \ref{LN3W2a} shows the behavior of
$\mathcal{N}_{\textrm{R}\textrm{S}}$ in terms of $u_1$ and $u_2$.
We see that if the accelerations of Rob and Steven satisfy
(\ref{root}), the entanglement between them will be removed,
completely. This is a remarkable result and it seems that this
contradicts the general behavior of fermionic entanglements under
the Unruh effect. However, $\mathcal{N}_{\textrm{R}\textrm{S}}$
represents the residual entanglement of Rob and Steven parties
after tracing over the Alice states. Thus, the entanglement level
for this system is lower than the entanglement of the whole
tripartite system, as Fig. \ref{LN3W2a} shows. Moreover, both Rob
and Steven are accelerated observers and so the rate of
entanglement degradation is such that the entanglement descends to
zero for finite accelerations.

In Fig. (\ref{LN3W2b}) we have plotted
$\mathcal{N}_{\textrm{A}\textrm{R}}$
($\mathcal{N}_{\textrm{A}\textrm{S}}$) which depends only on $u_1$
($u_2$). Since the entanglement is destroyed upon tracing over
Steven (Rob), the curve starts with $0.5$ at $u_1=0$ ($u_2=0$), as
for RS subsystem. But, here only one of the observers accelerates
and the entanglement degradation is not enough for vanishing the
entanglement at a finite or even infinite acceleration.

It must be noted that after tracing out any part of the tripartite
state (\ref{WRind}) we obtain a bipartite subsystem having some
residual entanglement, in an apparent contrast to the GHZ case.
One may say that the Unruh effect does not change the class of the
W state (\ref{Wket}). Notice that a W state remains entangled
after tracing out one of its parts. However, it is remarkable that
for RS subsystem the residual entanglement can completely be
removed for appropriate accelerations of Rob and Steven.

\section{Bosonic entanglements }

In this section we are going to discuss the effect of Unruh
temperature on the bosonic tripartite GHZ and W entanglements.
Because of the nature of bosonic states and hence the form of
corresponding Bogoliubov transformation (\ref{BosonBog}), our
calculations will be more complicated than the calculations for
the GHZ case discussed in the before section. We begin with the
bosonic GHZ state.

\subsection{The GHZ state}

Again we consider the GHZ state (\ref{GHZket}) which is written in
terms of  Minkowski modes. In the present argument we expand
Minkowski states $|0_{k_\textrm{R}}\rangle^+$ and
$|0_{k_\textrm{S}}\rangle^+$ in terms of Rindler bosonic states
for Rob and Steven, using the Bogoliubov transformation
(\ref{BosonBog}). Then we can write
\begin{eqnarray}\label{BGHZRin1}
        \ket{\textrm{GHZ}}_{\textrm{ARS}}&=&
        \frac{1}{\sqrt2}\left[
       \ket{0}
       \left(\frac{1}{\cosh{r_1}}\sum_{n}^{\infty}\tanh^n{r_1}\ket{n}_{\textrm{I}}\ket{n}_{\textrm{II}}\right)\left(
       \frac{1}{\cosh{r_2}}\sum_{m}^{\infty}\tanh^m{r_2}\ket{m}_{\textrm{I}'}\ket{n}_{\textrm{II}'}
       \right)\right.\\ \no
       &&\left.+\ket{1}\left(\frac{1}{\cosh^2{r_1}}\sum_{n=0}^{\infty}\tanh^n{r_1}\sqrt{n+1}
       \ket{n+1}_\textrm{I}\ket{n}_{\textrm{II}}\right)\right.\\
       &&\left.\times \left(\frac{1}{\cosh^2{r_2}}
       \sum_{m=0}^{\infty}\tanh^m{r_2}\sqrt{m+1}\ket{m+1}_{\textrm{I}'}\ket{m}_{\textrm{II}'}\right)\right].
\end{eqnarray}
The corresponding density operator
$\rho=|\textrm{GHZ}\rangle_{\textrm{ARS}}\langle \textrm{GHZ}|$
contains five partitions, however, as before we must trace out the
causally disconnected regions II and $\textrm{II}'$. Then we reach
to an infinite dimensional density matrix
\begin{equation}\label{BGHZRin2}
      \rho_{\textrm{A,I,I}'}=\frac{1}{2\cosh^2r_1\cosh^2r_2}\sum_{n,m}\tanh^{2n}r_1\tanh^{2m}r_2\,\rho_{nm},
\end{equation}
where
\begin{eqnarray}\label{rhonm}\nonumber
      \rho_{nm}&=&\ket{0,n,m}\bra{0,n,m}+
      \frac{\sqrt{n+1}}{\cosh^2{r_1}}\frac{\sqrt{m+1}}{\cosh^2r_2}\ket{0,n,m}\bra{1,
      n+1,m+1}+\frac{\sqrt{n+1}}{\cosh^2r_1}\frac{\sqrt{m+1}}{\cosh^2r_2}\\ \no
      &&\times\ket{1,
      n+1,m+1}\bra{0,n,m}+\frac{{n+1}}{\cosh^2r_1}\frac{{m+1}}{\cosh^2r_2}\ket{1,
      n+1,m+1}\ket{1,n+1,m+1}.
 \end{eqnarray}

\subsubsection{\textrm{A}-\textrm{RS}, \textrm{R}-\textrm{AS}, \textrm{S}-\textrm{AR} entanglements}

In order to quantify the entanglement of the ARS system described
by (\ref{BGHZRin2}), we invoke the logarithmic negativity
introduced in (\ref{LogNeg}). First we must calculate the
partially transposed density matrices
$\rho_{\widetilde{\textrm{A}},\textrm{I},\textrm{I}'}$,
$\rho_{\textrm{A},\widetilde{\textrm{I}},\textrm{I}'}$ and
$\rho_{\textrm{A},\textrm{I},\widetilde{\textrm{I}'}}$. These
matrices have infinite dimensions, however they are block diagonal
matrices. So, we encounter only square matrices located at each
block. For instance, the $(n,m)$ block of the Alice partially
transposed density matrix, is obtained as
\begin{equation}\label{BGHZPA}
       \left(\rho_{\widetilde{\textrm{A}},\textrm{I},\textrm{I}'}\right)_{nm}=\left(
       \begin{array}{cc} \tanh^2r_1 \tanh^2r_2&
       \frac{\sqrt{n+1}}{\cosh{r_1}}\frac{\sqrt{m+1}}{\cosh r_2}\\
       \frac{\sqrt{n+1}}{\cosh r_1}\frac{\sqrt{m+1}}{\cosh r_2}&
       \frac{n}{\sinh^2r_1}\frac{m}{\sinh^2r_2}
       \end{array}\right).
 \end{equation}
There are similar expressions for partially transposed density
matrices for Rob and Steven. Let $N_{nm}$ be the negative
eigenvalue of each block, then, the negative eigenvalue for the
whole matrix is $N=\sum_{n,m=0}^{\infty}N_{nm}$ which can be used
in (\ref{LogNeg}) to get the logarithmic negativity. We obtain the
negative eigenvalue for the block (\ref{BGHZPA}) as
\begin{eqnarray}\nonumber\label{BGHZPANnm}
      N^{\textrm{A}-\textrm{RS}}_{nm} &=& \frac{\tanh ^{2n}r_1\tanh
      ^{2m}r_2}{4\cosh ^2r_{{1}}\cosh^2r_2}\left( \tanh^2r_1\tanh^2r_2+{\frac{nm}{\sinh
      ^2r_1\sinh ^2r_2}}\right. \\  && \left. -\sqrt{\left( \tanh ^2r_1\tanh ^2r_2
      +{\frac{nm}{\sinh ^2r_1\sinh^2r_2}}\right)^2+\,{\frac{4(n+m+1)}{\cosh^2r_1\cosh^2r_2}}}\right).
\end{eqnarray}
Also, for the S-AR system, the negative eigenvalue for the $(n,m)$
block is obtained as:
\begin{eqnarray}\label{BGHZPSNnm}
      N^{\textrm{S}-\textrm{AR}}_{nm}&=&\frac{\tanh ^{2n}r_{{1}}\tanh ^{2m}r_2}{4\cosh ^{2}r_1
      \cosh ^{2}r_2}\left( \tanh ^2r_2+{\frac{(n+1)m}{\sinh^2r_1
      \sinh ^{2}r_2}}\right.\\\nonumber
      &&\left.-\sqrt{\left( \tanh^2r_2+\frac{(n+1)m}{\sinh^2r_1\sinh^2r_2}\right)^2+\,{\frac{4(n+1)}{\cosh^2r_1
     \cosh^2r_2}}}\right).
\end{eqnarray}
It turns out that the negative eigenvalue for the $(n,m)$ block of
the R-AS system, can be obtained by interchanging  $r_2$ and $r_1$
in Eq.(\ref{BGHZPSNnm}).

\begin{figure}
\subfigure[]{
    \label{BGHZPA:subfig:a}   \begin{minipage}[b]{0.4\textwidth}
      \centering
      \includegraphics[width=6cm,height=6cm]{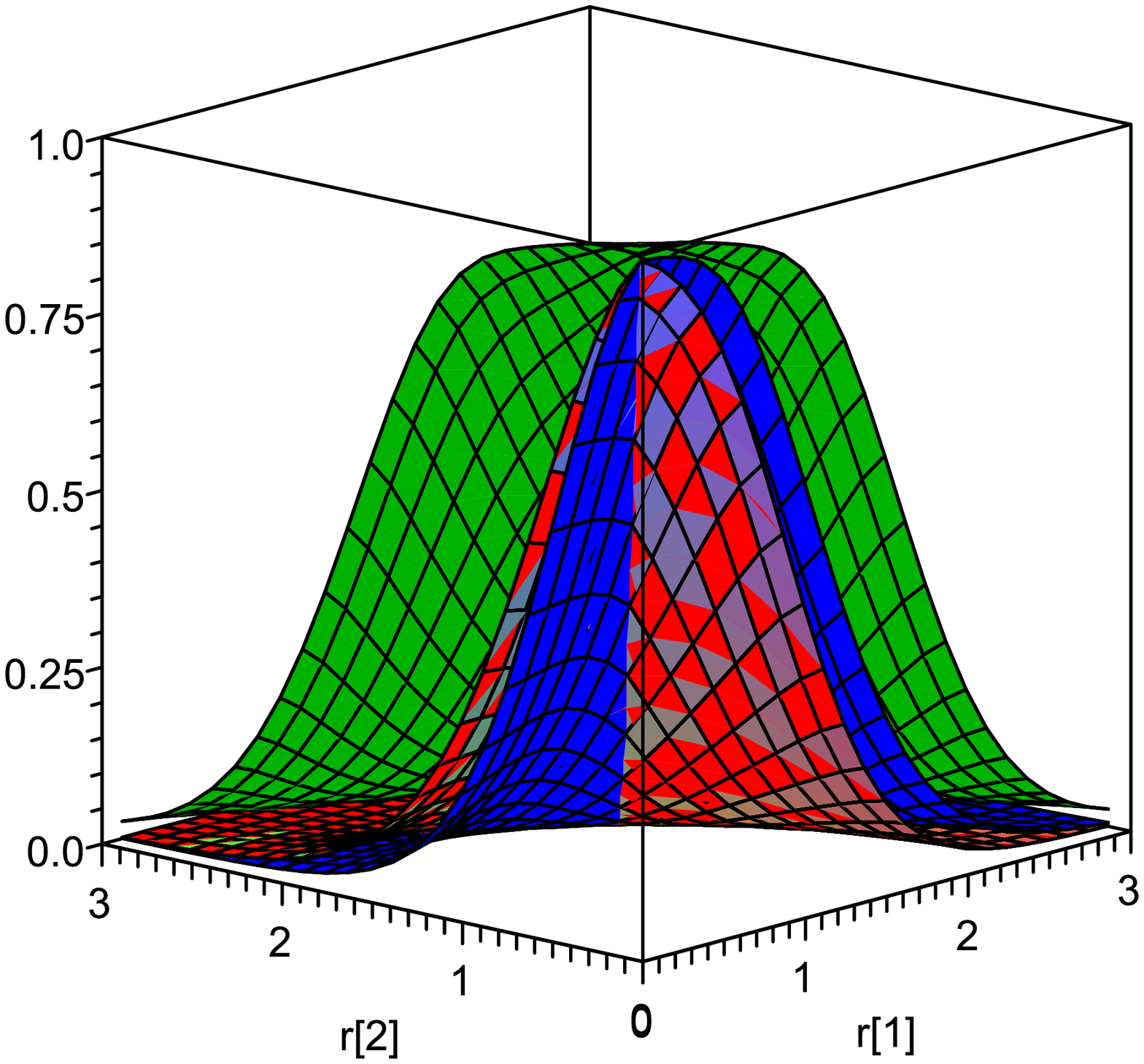}
    \end{minipage}}
\subfigure[]{
    \label{BGHZ3LN:subfig:b}     \begin{minipage}[b]{0.4\textwidth}
      \centering
      \includegraphics[width=6cm,height=6cm]{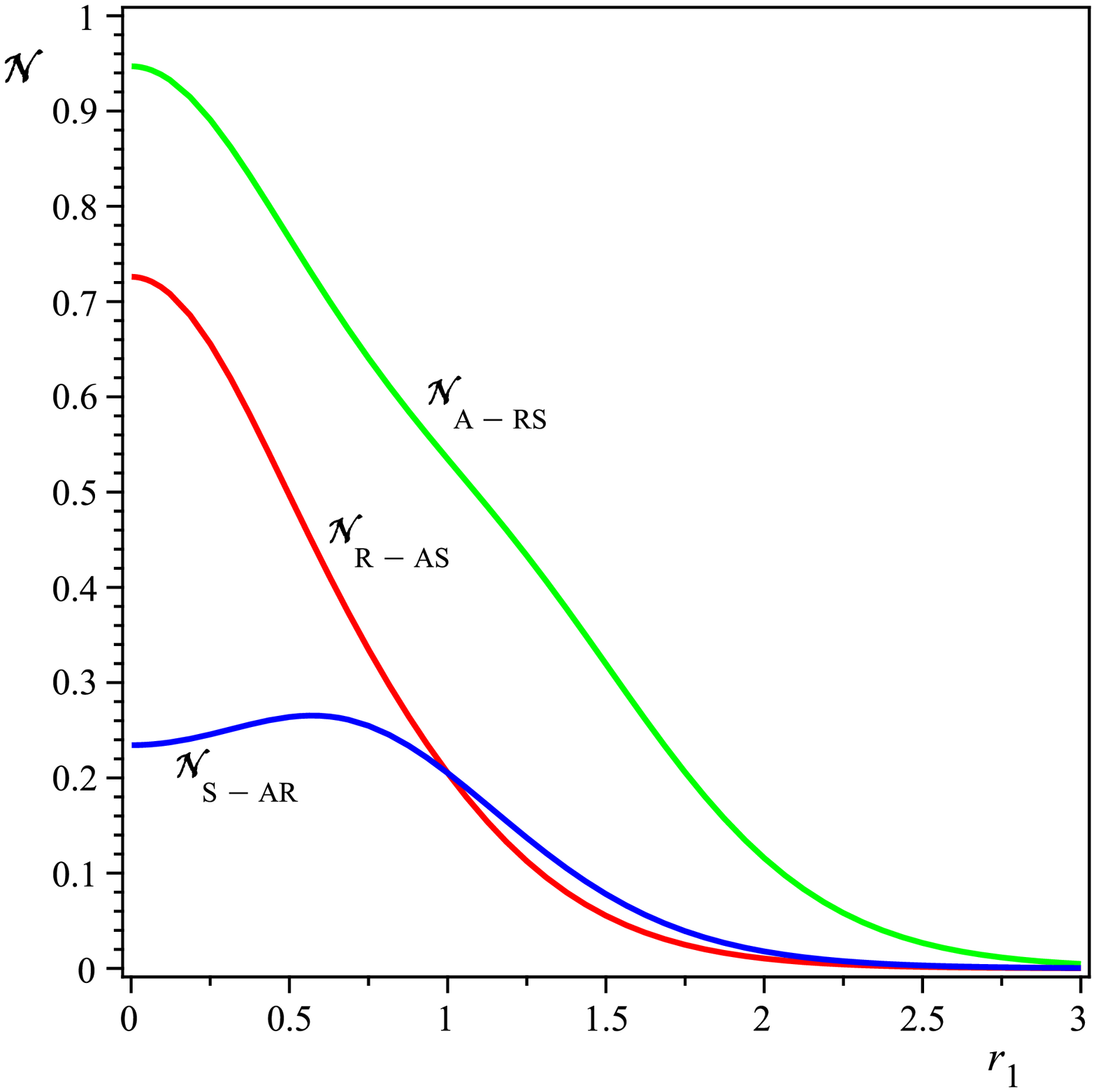}
    \end{minipage}}
\caption{(a) $\mathcal{N}_{\textrm{A}-\textrm{RS}}$,
$\mathcal{N}_{\textrm{R}-\textrm{AS}}$ and
$\mathcal{N}_{\textrm{S}-\textrm{AR}}$ for the bosonic GHZ state.
The surfaces $\mathcal{N}_{\textrm{R}-\textrm{AS}}$
  and $\mathcal{N}_{\textrm{S}-\textrm{AR}}$ have an intersection
  at $r_1=r_2$ and the surface
  $\mathcal{N}_{\textrm{A}-\textrm{RS}}$ covers them. Here the entanglement
asymptotically vanishes. (b) A section of figure (a) for a given
$r_2=1$. } \label{BGHZ3LN1}
\end{figure}

Now, the logarithmic negativity
$\mathcal{N}_{\textrm{A}-\textrm{RS}}$ becomes
\begin{equation}\label{LNA1}
      \mathcal{N}_{\textrm{A}-\textrm{RS}}=\log_{2}{\left(1-2\sum_{n,m=0}^{\infty}
      \mathrm{N}^{\textrm{A}-\textrm{RS}}_{nm}\right)},
\end{equation}
with similar expressions for
$\mathcal{N}_{\textrm{S}-\textrm{AR}}$ and
$\mathcal{N}_{\textrm{R}-\textrm{AS}}$. These functions are
plotted versus $r_1$ and $r_2$ in Fig. \ref{BGHZPA:subfig:a}. As
the figure shows, the surfaces
$\mathcal{N}_{\textrm{R}-\textrm{AS}}$ and
$\mathcal{N}_{\textrm{S}-\textrm{AR}}$ have an intersection at
$r_1=r_2$ and the surface $\mathcal{N}_{\textrm{A}-\textrm{RS}}$
covers them, as for the fermionic GHZ case. A section of these
surfaces for a given $r_2=1$ is plotted in Fig.
\ref{BGHZ3LN:subfig:b}. The distinction between the curves again
implies that in the considered tripartite system each part is
differently entangled to the other parts. In contrast to the
fermionic GHZ case, the logarithmic negativity in the present case
asymptotically vanishes.

\subsubsection{Entanglement of bipartite subsystems}

Entanglement of bipartite subsystems AR, AS and RS can be obtained
for each case by taking trace over the states in the otherwise
subsystem. After doing some manipulations it turns out that all
the resulting bipartite density matrices are diagonal and so, we
conclude that there is no entangled bipartite subsystem. This
resembles the fermionic GHZ state.

\subsection{THE  W state}

Now, let us apply the Bogoliubov transformation (\ref{BosonBog})
in the W state (\ref{Wket}) written in terms of Minkowski modes.
Then we reach to the following state,
\begin{eqnarray}\label{BWRin1}\no
       &&\ket{\textrm{W}}_{\textrm{ARS}}= \frac{1}{\sqrt3}\left[\ket{1}_M
       \left(\frac{1}{\cosh{r_1}}\sum_{n}^{\infty}\tanh^n{r_1}\ket{n}_{\textrm{I}}\ket{n}_{\textrm{II}}\right)\left(
       \frac{1}{\cosh{r_2}}\sum_{m}^{\infty}\tanh^m{r_2}\ket{m}_{\textrm{I}'}\ket{m}_{\textrm{II}'}
       \right)\right.\\
       && \left.
       +\ket{0}_{M}\left(\frac{1}{\cosh^2{r_1}}\sum_{n=0}^{\infty}\tanh^n{r_1}\sqrt{n+1}\ket{n+1}_\textrm{I}\ket{n}_{\textrm{II}}\right)
       \left(
       \frac{1}{\cosh{r_2}}\sum_{m}^{\infty}\tanh^m{r_2}\ket{m}_{\textrm{I}'}\ket{m}_{\textrm{II}'}
       \right) \right. \\ \no && \left.
       +\ket{0}_{M}\left(\frac{1}{\cosh{r_1}}\sum_{n=0}^{\infty}\tanh^n{r_1}\ket{n}_\textrm{I}\ket{n}_{\textrm{II}}\right)
       \left(
       \frac{1}{\cosh^2{r_2}}\sum_{m=0}^{\infty}\tanh^m{r_2}\sqrt{m+1}\ket{m+1}_{\textrm{I}'}\ket{m}_{\textrm{II}'}
       \right) \right].
\end{eqnarray}
Again tracing out the causally disconnected regions II and
$\textrm{II}'$, we reach to the following density matrix
\begin{equation}\label{BWRin2}
     \rho_{\textrm{A,I,I}'}=\frac{1}{{3}\cosh^2{r_1}\cosh^2{r_2}}\sum_{n,m=0}^{
     \infty}\tanh^{2n}{r_1}\tanh^{2m}{r_2}\,\rho_{nm}
\end{equation}
where
\begin{eqnarray}\label{Wrho1}
     \rho_{nm}&=&\ket{1nm}\bra{1nm}+\left(\frac{\sqrt{n+1}}{\cosh{r_1}}\ket{1nm}\bra{0,n+1,m}
     +H.C.\right)\\ \nonumber &&+\left(\frac{\sqrt{m+1}}{\cosh{r_2}}\ket{1nm}\bra{0,n, m+1}
     +H.C.\right)+
     \frac{n+1}{\cosh^2{r_1}}\ket{0,n+1,m}\bra{0,n+1,m}
     \\ \nonumber &&+\left(\frac{\sqrt{n+1}\sqrt{m+1}}{\cosh{r_1}\cosh{r_2}}\ket{0,n+1,m}\bra{0,n,m+1}+H.C.\right)
     +\frac{m+1}{\cosh^2{r_2}}\ket{0,n,m+1}\
     \bra{0,n,m+1}.
\end{eqnarray}
By inspection, we realize that the required partially transposed
density matrices deduced from (\ref{BWRin2}), are not block
diagonal. So the calculation of logarithmic negativity for these
density matrices by the trick of the previous subsection is
impossible and we encounter a complicated problem that can be
tackled by a numerical procedure. However, we do not follow this
here and content ourselves with an approximation valid only for
small $r_1$ and $r_2$. Thus, we can consider the summation
(\ref{BWRin2}) up to $m=n=1$, which leads to an $18\times 18$
matrix. Then, we see that the behavior of the entanglements in the
present case for small accelerations, is similar to what is shown
in Fig. \ref{BGHZ3LN1} for the GHZ state.

\subsubsection{Entanglement of bipartite subsystems}

\begin{figure}
\subfigure[]{
    \label{WNegA}     \begin{minipage}[b]{0.4\textwidth}
      \centering
      \includegraphics[width=6cm,height=6cm]{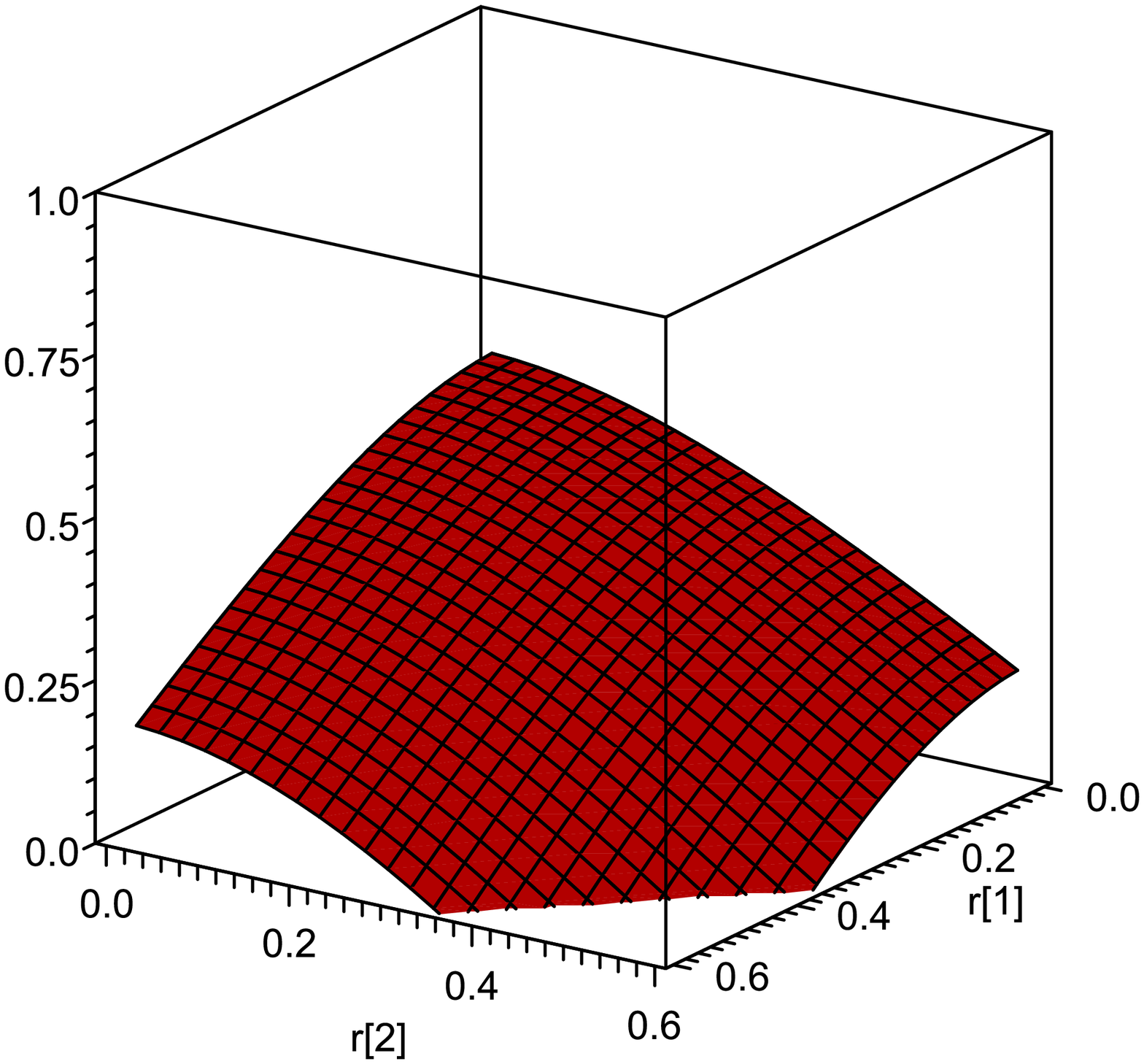}
    \end{minipage}}
\subfigure[]{
    \label{BWTA}     \begin{minipage}[b]{0.4\textwidth}
      \centering
      \includegraphics[width=6cm,height=6cm]{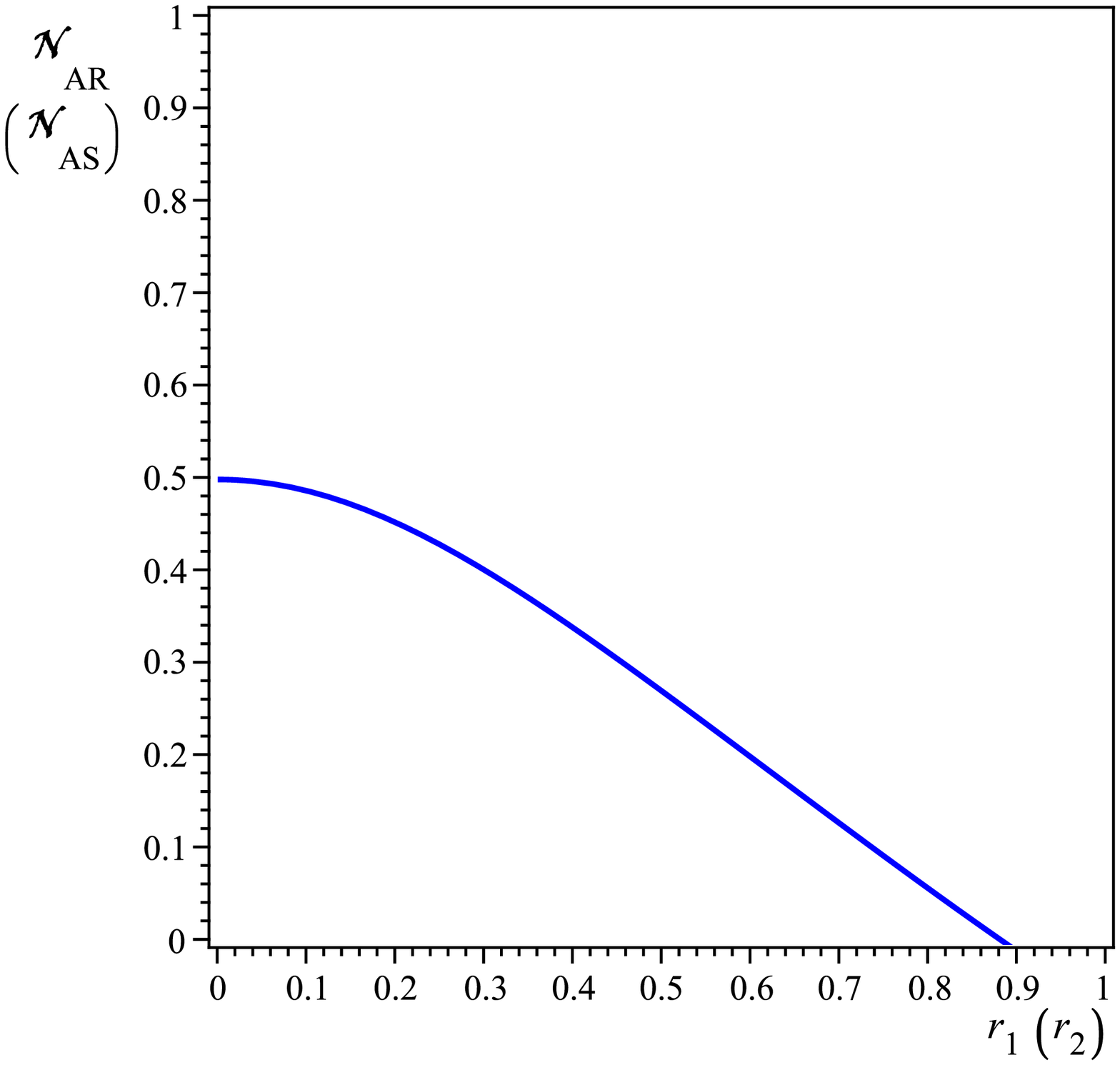}
    \end{minipage}}
\caption{The logarithmic negativity versus $r_1$ and $r_2$ for
bipartite subsystems when the tripartite system
     is in the bosonic W state.(a) The logarithmic negativity
$\mathcal{N}_{\textrm{R}\textrm{S}}$ vanishes for certain values
of $r_1$ and $r_2$ determined by the root of (\ref{WANeq}). (b)
The logarithmic Negativity $\mathcal{N}_{\textrm{A}\textrm{R}}$
($\mathcal{N}_{\textrm{A}\textrm{S}}$) vanishes at $r_1
(r_2)=\ln(1+\sqrt{2})$ corresponding to a certain value of Rob
(Steven) acceleration.} \label{WLNS}
\end{figure}

Let us trace out the Alice part of the state (\ref{BWRin2}). Then,
we reach to the following density matrix for the RS subsystem
\begin{equation}\label{WTA}
     \rho_{\textrm{I},\textrm{I}'}
     =\frac{1}{3\cosh^2{r_1}\cosh^2{r_2}}\sum_{n,m=0}^{\infty}\tanh^{2n}{r_1}\tanh^{2m}{r_2}\, \rho_{nm}
\end{equation}
where
\begin{eqnarray}\nonumber
      \rho_{nm}&=&\ket{nm}\bra{nm}+\frac{n+1}{\cosh^2{r_1}}\ket{n+1,m}\bra{n+1,m}
        +\frac{\sqrt{n+1}\sqrt{m+1}}{\cosh{r_1}\cosh{r_2}}\ket{1,n+1}\bra{1,m+1}\\
      &&+\frac{m+1}{\cosh^2{r_2}}
      \ket{n,m+1}\bra{n,m+1}.
\end{eqnarray}
After taking the partial transpose on Rob states we reach to a
block diagonal matrix with the following negative eigenvalue for
the $(m,n)$ block
\begin{equation}\label{WANeq}
      N^{\textrm{R}\textrm{S}}_{nm}=\frac{\tanh^{2n}{r_1} \tanh^{2m}{r_2}}{6\cosh^2{r_1}
      \cosh^2{r_2}
      }\left( a+b-\sqrt {
      \left( a+b \right) ^{2}-4\,ab+4\,\frac { \left( n+1 \right)  \left(
      m+1 \right) }{  \cosh^2{r_1}
      \cosh^2{r_2} }} \right)
\end{equation}
where
\begin{eqnarray}\label{WTNaA}\nonumber
      a&=&1+\frac {n}{
      \tanh^2{r_1} \cosh{r_1} }+\frac {m}{ \tanh^2{r_2}\cosh{r_2}},\\ \nonumber
       b&=& 2\ \tanh^2{r_1} \tanh^2{r_2}  +\frac { \left( n+1 \right)
      \tanh^2{r_2} }{\cosh^2{r_1}}+\frac { \left( m+1 \right)
      \tanh^2{r_1}}{\cosh^2{r_2}}.
\end{eqnarray}
The corresponding logarithmic negativity
$\mathcal{N}_{\textrm{R}\textrm{S}}$ is plotted in Fig.
\ref{WNegA}. As the figure shows, RS subsystem becomes
disentangled at certain finite values of $r_1$ and $r_2$. To
determine these values one should find the root of (\ref{WANeq}).

Now, tracing over the Steven states, we reach to the following
reduced density matrix for the AR subsystem
\begin{equation}\label{BWTS}
      \rho_\textrm{A,I}=\frac{1}{3\cosh^2{r_1}}\sum_{n=0}^{\infty}{\tanh^{2n}{r_1}\rho_n}
\end{equation}
where
\begin{eqnarray}
       \rho_{n}&=&
       \ket{1,n}\bra{1,n}+\left(\frac{\sqrt{n+1}}{\cosh{r_1}}\ket{1n}\bra{0,
       n+1}+H.C.\r \no \\ &&+\frac{n+1}{\cosh^2{r_1}}\ket{0, n+1}\bra{0,
       n+1}+\ket{0,n}\bra{0,n}.
\end{eqnarray}
After taking the partial transpose on the Alice states, we reach
to a block diagonal matrix that its $n$th block has the negativity
\begin{eqnarray}\label{BWTSNnm}
       N^{\textrm{A}\textrm{R}}_{n}&=&\frac{\tanh^{2n}r_1}{6\cosh^2r_1}
       \left( 1+ \frac {n}{\sinh^2r_1}+
       \tanh^2r_1\right.\\\nonumber
       &&\left.-\sqrt{\left(1+
       \frac {n}{\sinh^2r_1}+
       \tanh^2r_1\right)^2-4\,
       \tanh^2r_1+ \frac{4}{\cosh^2r_1}} \right),
\end{eqnarray}
which depends only on  $r_1$. Then, the logarithmic negativity is
obtained by
$\mathcal{N}_{\textrm{A}\textrm{R}}=\log_{2}{\left(1-2\sum_{n=1}^{\infty}
\mathrm{N}^{\textrm{A}\textrm{R}}_n\right)}$. It must be noted
that (\ref{BWTSNnm}) has a zero at $r_1=\ln(1+\sqrt{2})$,
independent of $n$. Then we have negativity only for $0\leq
r_1<\ln(1+\sqrt{2})$. The logarithmic negativity
$\mathcal{N}_{\textrm{A}\textrm{R}}$ is plotted in Fig.
\ref{BWTA}. As the figure shows, the entanglement vanishes at a
finite acceleration corresponding to  $r_1=\ln(1+\sqrt{2})$. It
can be shown that, the reduced density matrix for the AS subsystem
has the same form of (\ref{BWTS}), but $r_1$ is replaced by $r_2$.
So, the results indicated in Fig. \ref{BWTA} can also be
considered for the AS subsystem.

Again note that the entanglement level for each bipartite
subsystem is lower than that of whole system and here the Unruh
effect can completely remove the residual entanglements at finite
accelerations. Thus, for the bosonic W entanglement, all the
bipartite subsystems become disentangled at finite accelerations,
in contrast to the fermionic W states that its AR and AS
subsystems never become disentangled (see Fig. \ref{LN3W2}).

\section{conclusions}

Previously, in Ref. [15], the bosonic bipartite entanglement, and
in Ref.[16], the fermionic bipartite entanglement, was discussed.
However, in these works two observers were considered, one
inertial observer and one accelerated observer. However, our
setting is more general and so we obtain some new results that are
special to tripartite systems and distinguish this work from the
previous works.

In this work we considered the degradation of entanglement in
tripartite entangled states caused by the Unruh effect. In
particular, we considered two significant classes of tripartite
systems namely GHZ and W states. These entangled states were built
by three free modes of bosonic or fermionic quantum fields. One of
these modes was observed by an inertial observer Alice and the
other two modes were observed by uniformly accelerated observers
Rob and Steven. This leads to the detection of thermal radiation
by accelerating observers, which generally degrades the
entanglement in the system. We showed that the Unruh effect, even
for infinite accelerations, cannot completely remove the
entanglement in the fermionic GHZ and W states. On the other hand
for the bosonic states, we showed that the entanglement rapidly
drops and is erased for large values of accelerations.

We used the logarithmic negativity as a measure for these
tripartite entanglements, and interestingly, the logarithmic
negativity was not generally the same for different parts of the
system. This means that we encounter tripartite systems where each
part is differently entangled to the other two parts. For instance
in the fermionic or bosonic GHZ state, the Alice part is mostly
entangled to the Rob and Steven parts, for all accelerations. But
for W states this depends on the accelerations. Of course, for
determined accelerations it is possible that the entanglement be
the same for two parts of the system.

We also discussed the degradation of entanglement for bipartite
subsystems. Both for fermionic and bosonic GHZ states, tracing
over each part of the system leaves a disentangled bipartite
subsystem. However, tracing out any part of the fermionic or
bosonic W state leads to a bipartite system having some
accelerated-dependent entanglement. It was deduced that for the
fermionic W state, if the Alice part is traced out, the remaining
entanglement can vanish for certain finite accelerations. But, if
Rob or Steven part is traced out, the remaining entanglement will
decrease to nonzero values, asymptotically. For the bosonic W
state, we showed that all the bipartite entanglements can vanish
for determined finite accelerations.

\end{spacing}

\end{document}